\documentstyle[12pt]{article}

\def\8{\infty}
\def\oh{\frac{1}{2}}

\def\oq{\frac{1}{4}}

\def\d{\partial}
\def\i{\imath\,}
\def\undertext#1{\vtop{\hbox{#1}\kern 1pt \hrule}}
\def\ra{\rightarrow}

\def\Ra{\Rightarrow}
\def\lra{\longrightarrow}

\def\lrb#1{\left(#1\right)}
\def\O#1{O\left(#1\right)}
\def\VEV#1{\left\langle\,#1\,\right\rangle}
\def\tr{\hbox{tr}\,}
\def\trb#1{\tr\lrb{#1}}

\def\dbyd#1#2{\frac{d#1}{d#2}}
\def\pp#1{\frac{\partial}{\partial#1}}

\def\ff#1{\frac{\delta}{\delta#1}}
\def\fbyf#1#2{\frac{\delta#1}{\delta#2}}

\def\br{\\ \nonumber & &}
\def\inv#1{\frac{1}{#1}}
\def\be{\begin{equation}}
\def\ee{\end{equation}}
\def\bea{\begin{eqnarray} & &}
\def\eea{\end{eqnarray}}
\def\ct#1{\cite{#1}}
\def\rf#1{(\ref{#1})}
\def\EXP#1{\exp\left(#1\right)}

\def\MAT{{\it Mathematica }}
\def\LHS{left hand side}
\def\RHS{right hand side}
\def\MINT#1{\int \frac{d^d #1}{(2\pi)^d}}
\def\PDF{probability distribution function }

\def\SD{Schwinger-Dyson }
\def\TEXP#1{\,\hat{T}\exp\left(#1\right)}

\def\N#1{\nabla_{#1}}

\def\YM{Yang-Mills }

\def\X#1{\hat{X}_{#1}}
\def\D#1{\pp{\X{#1}}}
\def\U#1#2{\TEXP{\i\int_{#1}^{#2}\,d\,p_\mu\,\X{\mu}}}
\def\TRU{\tr \TEXP{\i\oint\,d\,p_\mu\,\X{\mu}}}
\def\W#1{\trb{\TEXP{\oint_{C_{#1}} A_{\mu} d x_{\mu}}}}

\begin{document}

\begin{titlepage}

{\bf \today}\hfill	  {\bf PUPT-1509}\\
\begin{center}

{\bf  SECOND QUANTIZATION OF THE WILSON LOOP  }

\vspace{1.5cm}

{\bf A.A.~Migdal}

\vspace{1.0cm}

{\it  Physics Department, Princeton University,\\
Jadwin Hall, Princeton, NJ 08544-1000.\\
E-mail: migdal@puhep1.princeton.edu}

\vspace{1.9cm}

\end{center}

\abstract{ Treating the QCD Wilson loop as amplitude for the propagation of the
first quantized particle we develop the second quantization of the same
propagation. The operator of the particle position $\hat{\cal X}_{\mu}$ (the
endpoint of the "open string") is introduced as a limit of the large $N$
Hermitean matrix. We then derive the set of equations for the expectation
values of the vertex operators $\VEV{ V(k_1)\dots V(k_n)} $. The remarkable
property of these equations is that they can be expanded at small momenta (less
than the QCD mass scale), and solved for expansion coefficients. This provides
the relations for multiple commutators  of position operator, which can be used
to construct this operator. We employ the noncommutative probability theory and
find the expansion of the operator $\hat{\cal X}_\mu $ in terms of products of
creation operators $ a_\mu^{\dagger}$. In general, there are some free
parameters left in this expansion. In two dimensions we fix parameters uniquely
from the symplectic invariance. The Fock space of our theory is much smaller
than that of perturbative QCD, where the creation and annihilation operators
were labelled by continuous momenta. In our case this is a space generated by
$d = 4$ creation operators. The corresponding states are given by all sentences
made of the four letter words. We discuss the implication of this construction
for the mass spectra of mesons and glueballs.
}
\vfill
\end{titlepage}
\tableofcontents

\section{Introduction}

The interpretation of the Wilson loop $W(C)$ as an amplitude of the charge
propagation has a long history. Already in early Feynman's work the abelian
phase factor was interpreted this way. Wilson elaborated on that in his famous
paper "Confinement of Quarks", taking the first quantized picture.

The quark propagator in this picture is given by the sum over closed loops $ C$
of the loop amplitudes $ \sum_C W(C) G(C) $, where $ G(C) $ is the amplitude
for the free quark. In the large $N$ limit  there are no dynamical quarks,
which makes this picture especially attractive. There is a tempting analogy
with the string theory. In fact, the string theory originated as the planar
diagram theory of quark confinement and was boosted by the 't Hooft's discovery
\ct{Hooft} of the large $N$ limit.

Now, what would be the second quantized picture of this process? This is not
just the matter of mathematical curiosity. The development of Feynman-Wilson
picture for the last 20 years led us to the dead end. We derived\ct{MM}
nonlinear loop equation for $W(C)$, but this equation turned out to be too hard
to solve.

In two dimensions, where there are  only global degrees of freedom, related to
topology of the loop, the loop equations were solved\ct{KK}. In four dimensions
the best bet was the Eguchi-Kawai reduction, which, unfortunately worked only
on the lattice. Taking the continuum limit in the EK model appears to be as
hard as in the original partition function of the lattice theory. This forces
us to look for the new approaches, such as the second quantized picture.

\subsection{Classical vs Quantum Loops}

The second quantization of the usual field theory involves the following steps
\be
fields \Ra operators \Ra commutators \Ra H \Psi = E \Psi.
\ee
We would like to follow these steps in QCD for the particle position field $
\X{} $. The parametric invariance of the Wilson loop would make this one
dimensional topological field theory. The Hamiltonian $H$ vanishes in such
theory, as the field operators do not depend of the (proper) time.

To understand this striking phenomenon, let us consider the following simple
example:  Wilson loop in constant abelian electromagnetic field $ F_{\mu\nu} $
\be
W(C) = \EXP{ \oh \i F_{\mu\nu} \, \oint_C x_\mu \, d x_\nu}.
\ee
The exponential here represents symplectic form which immediately leads to the
commutation relations
\be
[\X{\mu},\X{\nu}] = \i\,(F^{-1})_{\mu\nu}.
\ee
of $\oh d $ harmonic oscillators.

This corresponds to the following momentum space path integral\footnote{Later
we
shall slightly modify this formula, to take into account the discontinuities
of the local momentum $p(s)$.}
\be
W(C) \propto \inv{N} \int D p(.) \delta^d(p(s_0)) \, \EXP{\i \oint d x_\mu
p_\mu}\TRU,
\label{ANZATZ}
\ee
One may regard  $p'_\mu(s) $ as Lagrange multiplier for the constraint on the
coordinate $ x_\mu(s) \approx \X{\mu} $ modulo total translations $ \delta
x_\mu(s) = const $. The constraint
\be
p(s_0) = 0
\ee
is needed to eliminate the translations of $p(s)$. The choice of $ s_0$ is
arbitrary, it is a matter of convenience.

The trace is calculable for the oscillator, it yields
\be
\TRU \propto \EXP{ - \oh \i (F^{-1})_{\mu\nu}\, \oint  p_\mu d p_\nu }.
\ee
The integration over $ p(.) $ correctly reproduces the initial formula for
$W(C)$. For the readers reference, this functional integral is computed in
Appendix A, including the normalization (irrelevant for our present purposes).

Note, that the operators $ \X{\mu} $ do not commute, and, hence, cannot be
simultaneously diagonalized. So, we cannot integrate out the Lagrange
multiplier and claim that $ x_\mu(s) $ coincides with one of the eigenvalues
of $\X{\mu}$.

The Wilson loop enters the physical observables inside the path integral, such
as
\be
\int_0^\8 d T \int D x(.) \delta^d(x(s_0))\EXP{- \int_0^T d t \lrb{ \oq
\dot{x}^2 + m^2}} W(C)
\ee
for the scalar particle propagator.
With our representation after integrating out $x(.)$ this becomes
\be
\int_0^\8 d T \int  D p(.) \TEXP{\int_0^T d t\lrb{ \i\, \dot{p}_\mu \X{\mu}-
H(p)}}
\ee
with the (proper time) Hamiltonian
\be
H = m^2 + p_\mu^2.
\ee
For the Dirac particle we get the same formula with the Dirac Hamiltonian
\be
\label{HDIR}
H = m + \i \gamma_\mu p_\mu
\ee
The only difference with the usual Hamiltonian dynamics is the replacement of
the particle position $x(t)$ by the operator$\X{}$ in the $ \dot{p}x $ term.
The implications of this replacements will be studied later.

\subsection{Master Field and Eguchi-Kawai Reduction}

In a way, our approach is an implementation of the Witten's master field
idea\ct{Witten}. He suggested to look for the classical $x$-independent field
$\hat{A}_\mu$ in coordinate representation
\be
\label{MASTER}
W(C) = \inv{N} \tr\TEXP{\oint_C d x_\mu \hat{A}_\mu}.
\ee
Clearly, the naive implementation, with $\hat{A}_\mu$ satisfying the \YM
equations, does not work. There are corrections from the fluctuations of the
gauge field. These corrections are taken into account in the loop
equations\ct{MM,Mig83}, so it is worth trying to find effective master field
equation for $\hat{A}_\mu$ from the loop equations.

In some broad sense the master field always exists. It merely represents the
covariant derivative operator
\be
\N{\mu} = -\i {\cal P}_\mu + {\cal A}_{\mu}(0),
\ee
where ${\cal P}_\mu $ is momentum operator and ${\cal A}_\mu$ is the gauge
field operator in the Hilbert space. As one may readily check, the matrix trace
of the ordered exponential of this operator reduces to the operator version of
the loop exponential times the unit operator in Hilbert space
\bea
\tr \TEXP{\int_{0}^{0} \N{\mu} d x_\mu} = \tr \TEXP{\int_{0}^{0} {\cal
A}_\mu(x) d x_\mu},\br
{\cal A}_\mu(x) = \EXP{-\i {\cal P}_\nu x_\nu} {\cal A}_\mu(0)\EXP{\i {\cal
P}_\nu x_\nu}.
\eea

Taking the  trace of this relation in Hilbert space of large $N$ QCD, and
dividing by trace of unity, we would get the usual Wilson loop. We could avoid
division by an infinite factor by taking the vacuum average instead.
\be
\left\langle 0 \right | \TEXP{\int_{0}^{0} \N{\mu} d x_\mu} \left | 0 \right
\rangle = \inv{N}\,\tr \TEXP{\int_{0}^{0}  A_\mu(x) d x_\mu}
\ee
This formal argument does not tell us how to compute this master field. The
operator version of the \YM equations is not very helpful.
It reads
\be
[\N{\mu},[\N{\mu},\N{\nu}]] = g_{eff}^2 \pp{\N{\mu}}.
\ee
The operator $ \pp{\N{\mu}} $ in turn, is given by its commutator relations
with $\N{\mu}$.

Formally, inserting the left side of this equation inside the trace in the
master field Anzatz \rf{MASTER}  we would reproduce the correct loop equation
provided the trace of the open loop is proportional to the delta function
$\delta^d(x-y)$. The normalization would  come out wrong, though, unless we
include the divergent constant $\delta^d(0) = \Lambda^d$ in effective coupling
constant
\be
g_{eff}^2 = g_0^2 \Lambda^d
\ee

The Eguchi-Kawai reduction\ct{EK} was, in fact, a realization of this scheme.
There, the large $N$ matrix $\hat{A}_\mu$ represented the covariant derivative
operator in above sense, except it was not a single matrix. Its diagonal
components played the role of momenta $\hat{A}_\mu^{ii} = -\i p_\mu^i $. These
were classical, or, to be more precise, quenched\ct{Quench} momenta.

The off-diagonal components were fluctuating with the weight $
\EXP{-\Lambda^d\, L_{YM}}$, corresponding to the \YM Lagrangean for the
constant matrix field. The cutoff $\Lambda$ explicitly entered these formulas.
In effect, one could not take the local limit in the quenched EK model.
Apparently something went wrong with implementation of the master field.

\subsection{Noncommutative Probability Theory}

Recently the large $N$ methodology was enriched\ct{Douglas94,Gross94} by using
so called  noncommutative probability theory\ct{Voiculesku}.
In particular, in the last paper\ct{Gross94} the master field was constructed
by similar method in the QCD2.

This amounts to the deformation of the commutation relation
\be
a_\mu(p) a_\nu^{\dagger}(p') - q \, a_\nu^{\dagger}(p')a_\mu(p) =
\delta_{\mu\nu}\delta^d(p-p')
\ee
Instead of the usual Bose commutation relations, with $ q = 1 $ one should take
$q = 0 $, which makes it  so called Cuntz algebra. This algebra in the context
of the planar diagram theory in QCD was first discovered long ago by Cvitanovic
et.al.\ct{Cvitanovic}.

As noted in\ct{Gross94}, this algebra corresponds to Boltzmann statistics. The
origin of the Boltzmann statistics in the large $N$ limit is clear: at large
rank each component of the matrix can be treated as unique object, neglecting
the probability of the coincidences of indexes. The indexes could be dropped
after this.

This interesting construction proves the concept, but we cannot be completely
satisfied with the fact that the operator representation in this paper is gauge
dependent. It used the fact that in the axial gauge there is no interaction in
QCD2, so that the only deviation from the free field behaviour of the Wilson
loop comes from the noncommutativity of the (Gaussian) gauge potential at
various points in space.

This theory is purely kinematical, it does not make use of the large $N$
dynamics. The Fock space is only a slight modification of the perturbative
Fock space. The color confinement was not taken in consideration. There should
exist a much tighter construction, built around the nonperturbative Fock space.

Also, the gauge dependence should be absent in the final formulation of the
theory. This resembles the early attempts to solve the Quantum Gravity in 2D by
means of perturbative Liouville theory. Taking the conformal gauge and using
the methods of conformal field theory, one was forced to work in the space of
free 2D particles,  with the Bose operators $a_n, a^{\dagger}_n$.

However, as we know now, the Fock space of 2D Gravity is much smaller. It is
described by the matrix models (or topological field theories), with only {\em
discrete} states. If we now employ the same noncommutative probability theory,
we could reinterpret the matrix models of 2D Gravity in terms of a single pair
of operators $a,a^{\dagger}$.

To conclude this discussion, the ideas of the noncommutative probability are
quite appealing, but implementation of these ideas in QCD is not yet finished.
The correct nonperturbative Fock space in confining phase of QCD is yet to be
found.

\subsection{Position Master Field}

In the present work, we follow the same general large $N$ philosophy, but
interpret the  master field as the momentum operator $ \hat{P}_\mu = \i
\hat{A}_\mu$ of the endpoint of the QCD string. With the second quantization,
developed below, we rather construct the position operator $\X{}$. After that
we shall reconstruct the momentum operator as well.

The advantage of the position operator is that (up to irrelevant global
translational mode) we expect its spectrum to be bounded from above in
confining phase.

The commutators $[\X{\mu},\X{\nu}] $ describe the "area inside the Wilson
loop". We could work with expectation values with products of such operators.
As for the momentum operators, the  high momentum region corresponds to small
distance singularities of perturbative QCD. It  makes the matrix elements of
commutators of the Witten's master field infinite.

To rephrase it, the Wilson loop in coordinate space is singular. It cannot be
expanded in powers of the size of the loop. There are the gluon exchange
graphs, which lead to singularities at intersections and corners. The gluon
exchange graphs for $W(C)$ scale as powers of $|C|^{4-d} $.

In two dimensions the singularities reduce to existence of various topological
sectors, with different Taylor expansions. Without self-intersections, there is
a pure area law, with intersections there are polynomials of areas inside
petals times exponentials\ct{KK}.

In four dimensions there are logarithms, corresponding to the asymptotic
freedom. It is not clear how  to reproduce these properties by any constant
large $N$ matrix $\hat{A}_\mu$.

With the position operator $\X{}$, as we shall see, the correct coordinate
dependence is built in our Anzatz \rf{ANZATZ}. In particular, the $\delta$
function in the loop equation comes out automatically, without any assumptions
about spectrum of $\X{}$. We escape the quenched EK model paradox of extra
factor $\Lambda^d$ in effective coupling constant.

\section{Loop Equations}

In this section we review and further develop the method of the loop equations
in the large $N$ QCD, first in coordinate space, then in momentum space.

\subsection{Spatial Loops}

 We describe these loops by a periodic function of a proper time $ s $
\be
C:  x_{\mu}(s)\;, x_{\mu}(s+1) = x_{\mu}(s).
\ee
The following \SD equation is to be used
\bea
0 = \int D A \,\tr \int_0^1 d t_1 x_{\nu}'(t_1)\,
g_0^2\,\ff{A_{\nu}(x(t_1))}\br
\EXP{ \int d^d x \frac{\tr F_{\mu\nu}^2}{4 g_0^2}}\, \hat{T} \EXP{\int_0^1\,d s
A_{\mu}(x(s))x'_{\mu}(s)}.
\eea
The variation of the first exponential yields the classical part of equation of
motion,
$$
 \left[D_{\mu}\,F_{\mu\nu}\right] ,
$$
whereas the second (ordered) exponential provides the commutator terms, present
only for the self-intersecting loops. \footnote{These self-intersections are
responsible for all the "quantum corrections" to the classical equations of
motion. Dropping them as "rare" or "exceptional" configurations would bring us
back to the classical Yang-Mills theory.}

These terms involve $ g_0^2 \,\oint d t_1 \oint d t_2
\delta^d\left(x(t_1)-x(t_2)\right)\delta_{\mu\nu} $ times the product of two
ordered exponentials,   corresponding to the two petals  $ C_{12}, C_{21} $ of
our  loop $C$. The $ \delta $ function ensures that both these petals are
closed loops, as it is required by the gauge invariance.

\input epsf \centerline{ \epsfbox{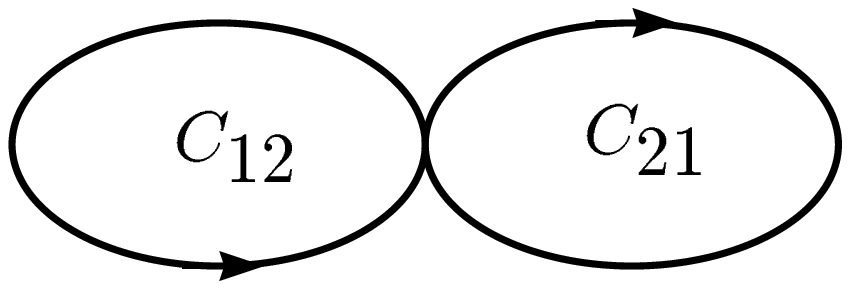}}

The classical part can be represented\ct{Mig83,Pol87} as the functional Laplace
(or Levy) operator
\be
\hat{L} =  \int_{0}^1 d t_1\, \ff{x_{\mu}(t_1)}\int_{t_1-0}^{t_1+0} d t_2 \,
\ff{x_{\mu}(t_2)}
\ee
acting on the loop functional
\be
W[C] = \inv{N} \VEV{\tr \hat{T} \EXP{ \int_0^1 d s A_{\mu}(x(s)) x'_{\mu}(s)}}.
\ee
This Levy operator is the only parametric invariant second order operator, made
of functional derivatives. It picks up the contact term in the second
functional derivative
\bea
\fbyf{W[C]}{x_{\mu}(t_1) \delta x_{\mu}(t_2)} \propto \br
\delta(t_1-t_2) x'_{\nu}(t_1)\,\inv{N} \tr \hat{T}
\left[D_{\mu}\,F_{\mu\nu}(x(t_1))\right] \,\EXP{ \int_0^T A_{\mu}(x(s))
x'_{\mu}(s)} \br
+ regular \, terms.
\eea
On the right hand side, the average product of two loop functionals reduces at
large $N$ to the product of averages
\bea
\inv{N} \tr \hat{T} \EXP{ \int_{t_1}^{t_2}d s_1 A_{\mu}(x(s_1)) x'_{\mu}(s_1)}
\,\br
\inv{N} \tr \hat{T} \EXP{ \int_{t_2}^{t_1+1} d s_2 A_{\mu}(x(s_2))
x'_{\mu}(s_2)} \br
\stackrel{N\ra\8}{\lra} W[C_{12}] W[C_{21}]
\eea
We thus arrive at the conventional loop equation
\be
\hat{L} W[C] = N g_0^2 \oint d t_1 x'_\mu(t_1) \int_{t_1^+}^{t_1^-} dt_2
x'_\mu(t_2) \delta^d\left(x(t_1)-x(t_2)\right) W[C_{12}] W[C_{21}],
\ee
where the second integral goes around the loop avoiding the first point $t_1$.
Thus, the trivial self-intersection point at $t_2 = t_1$ is eliminated
(see\ct{Mig83} for details).

Singularities in this  equation should be properly understood. The double
integral of the $d$ dimensional $ \delta $ function does not represent an
ordinary number in more than two dimensions. This is a singular functional in
the loop  space, the only meaning of which could be provided by an integration
with some weight function.

Otherwise we should mess up with the gauge invariant regularization of the
delta function with inevitable ambiguity of the point splitting procedure. One
such regularization is given by the lattice gauge theory, but the loss of the
space symmetry makes the lattice loop equation both ugly and useless. Even the
beautiful Eguchi-Kawai reduction could not save the lattice loop equation. It
did not offer any alternatives to the lattice gauge theory.

\subsection{Momentum Loops}

Let us now forget about all the lattice artifacts and come back to continuum
theory. The most natural thing to do with the delta function is to switch to
momentum space. The parametric invariant functional Fourier transform of the
Wilson loop is defined as follows
\be
W[p] = \int  D C W[C]\EXP{-\i\int_0^1 d s p_{\mu}(s)x'_{\mu}(s) },
\label{WPX}
\ee
where
\be
D C = \left(\prod_{s=0}^{s<1} d^d x(s)\right) \, \delta^d(x(s_0)).
\ee
The $\delta $ function here fixes one arbitrary point $ x(s_0)= 0 $ at the
loop, the choice of $ s_0 $ does not influence observables in virtue of
translational invariance. Note, that the mass center fixing condition $
\int_0^1 ds x(s) =0 $ is not parametric invariant, and would require the
Jacobian. This is not so difficult to do, but our choice of one point fixing is
simpler.

As it was noted in\ct{Mig86} this measure factorizes for self intersecting
loops on the right hand side of the loop equation. The argument goes as follows
(assuming $ t_1 < t_2 $, and choosing $ s_0 = t_1 $)
\bea
\left(\prod_{s=0}^{s<1} d^d x(s)\right)\, \delta^d(x(t_1)) \, \delta^d
\left(x(t_1)-x(t_2)\right)= \br
\left(\prod_{\tau=t_1}^{\tau<t_2} d^d x(\tau) \right)\, \delta^d(x(t_1)) \,
\left(\prod_{\sigma=t_2}^{\sigma < t_1+1} d^d x(\sigma)\right)\,
\delta^d(x(t_2)) = D C_{12} \, D C_{21}.
\eea
The $ \delta $ function for the loop self-intersection, $ \delta^d
\left(x(t_1)-x(t_2)\right) $ was combined with the point fixing $ \delta $
function $ \delta^d(x(t_1)) $ to provide the point fixing $ \delta $ functions
for both petals. Note that this fits nicely large $N$ factorization of the
Wilson loops. For the abelian theory such factorization of the measure would be
useless.

The functional Laplace operator on the left hand side of the loop equation
transforms to the momentum space as follows
\be
\int_0^1 d t_1  \ff{x_{\mu}(t_1)}\int_{t_1-0}^{t_1+0} d t_2 \ff{x_{\mu}(t_2)}
\Ra
-\int_0^1 d t_1 p'_{\mu}(t_1) \int_{t_1-0}^{t_1+0} d t_2 p'_{\mu}(t_2) .
\ee
As was discussed at length in\ct{Mig86}, for continuous
momentum loops this integral vanishes, but for the relevant loops with
finite discontinuities, $ \Delta p_k = p(t_k+0) - p(t_k-0) $, the Laplace
operator reduces to the sum of their squares $ \sum_k \left(\Delta
p_k\right)^2 $. These discontinuities are induced by emission and absorption of
gluons. We shall discuss this relation in more detail later.

In virtue of periodicity and symmetry  this can also be written as the {\em
ordered} integral
\bea
\int_0^1 d t_1 p'_{\mu}(t_1) \int_{t_1-0}^{t_1+0} d t_2 p'_{\mu}(t_2)=\br
-\int_0^1 d t_1 p'_{\mu}(t_1) \int_{t_1}^{t_1+1} d t_2 p'_{\mu}(t_2)=\br
-\int_0^1 d t_1 p'_{\mu}(t_1) \int_{t_1}^{1} d t_2 p'_{\mu}(t_2)\br
-\int_0^1 d t_1 p'_{\mu}(t_1) \int_{0}^{t_1} d t_2 p'_{\mu}(t_2)=\br
-2 \, \int_0^1 d t_2 p'_{\mu}(t_2) \int_{0}^{t_2} d t_1 p'_{\mu}(t_1)
\eea

We should order the integration variables on the right as well, to avoid the
trivial point $ t_1 = t_2 $ (one could trace this prescription back to the
definition of the ordered exponent as an infinite product of the {\em linear}
link factors $ 1 + A_\mu(x) d x_\mu $, which can be varied only once with
respect to $ A_\mu(x)$). After all this, we write the momentum loop equation as
follows
\be
W[p]\int_0^1 d t_2 p'_{\mu}(t_2) \int_0^{t_2} d t_1 p'_{\mu}(t_1)
=  \int_0^1 d t_2 \ff{p_{\mu}(t_2)} \int_0^{t_2} d t_1 \ff{p_{\mu}(t_1)}
W\left[p_{12}\right] W\left[p_{21}\right].
\ee
The parts $ p_{12}, p_{21} $ of original momentum loop are parametrized
by smaller proper time intervals $ T_{12} = t_2-t_1, T_{21} = 1+t_1-t_2 $. This
can be rescaled to $ (0,1) $ interval in virtue of the parametric invariance.
With the manifestly invariant Anzatz \rf{ANZATZ}, such rescaling is
unnecessary.

Note that the momentum loops, unlike the coordinate loops, are not closed.
There are inevitable discontinuities, in particular, both parts break  at the
matching point. Even if we would have assumed the initial loop $ p(s) $ to be
continuous, the $W$ functionals on the right hand side would enter with the
broken loops. Therefore, we have to allow such discontinuities on the left side
as well, to get the closed equation.

The normalization of the momentum loop functional was chosen to absorb the bare
coupling constant. The original normalization $ W[C]_{x(.)=0} = 1 $ results in
the relation for the bare coupling constant
\be
 N g_0^2 = \int Dp W[p]
\ee
(with the same measure $Dp$ as for the coordinate loop $C$).

We could treat this relation as a definition of the bare coupling constant.
Here is where the ultraviolet divergences show up. We should cut these
momenta, say, by the Gaussian cutoff $ \EXP{-\epsilon \int_0^1 d s p^2(s)} $.
Surprisingly, such a brute force regularization preserves the gauge
invariance, as our momenta were not shifted by a gauge field (see
\ct{Mig86}).

This resembles the solution of $ O(N)$  $\sigma$ model in two dimensions, where
the $ \VEV{n_i n_i} $ propagator was found to be the free one, with the mass
given by the expectation value $\VEV{\sigma} $ of Lagrange multiplier for the
constraint $n_i n_i =1$. This expectation value was then determined by the
equation
$$
1 =\VEV{n_i n_i} = N g_0^2  \int \frac{d^2 k}{4 \pi^2} \inv{k^2 + \VEV{\sigma}}
$$

In both cases, one could regard this normalization condition as definition of
unobservable bare coupling constant in terms of the ultraviolet cutoff. For the
observables one could just leave the physical mass scale as the adjustment
parameter.

The possibility to eliminate the bare coupling constant plus the scale
invariance of the loop equation $ W[p] \Ra \alpha^{-4} \, W[\alpha \,p] $ makes
this equation manifestly renormalizable.  The physical mass scale cannot be
found from the loop equation, it should rather be fitted to experiment. Then,
just for the internal consistency check, the bare coupling constant can be {\em
computed} as a function of the cutoff and the physical  mass.

\subsection{Planar Graphs in Momentum Loop Space}

This is not to say, that the perturbation theory is unrecoverable from the
momentum loop equation. The reader is invited to follow the direct derivation
of the Faddeev-Popov diagrams from this equation\ct{Mig86,BVM}, starting with
\be
W_0[p] =  N g_0^2 \MINT{q}\delta[p(.)-q]
\label{W0}
\ee
where
\be
\delta[p(.)-q] \equiv \prod_{s=0}^1\delta^d(p(s)-q)
\ee
is the functional $ \delta $ function.

In higher orders there are higher functional derivatives of the functional
delta functions $ \ff{p(\tau_m)} \delta[p(.)-q_l] $.  The integration variables
$ q_l$ serve as the momenta inside the loops of planar graphs (see\ct{BVM} for
details).

Say, in the second order we would have two momenta $q_1, q_2$ in two windows of
the one-gluon graph on a disk. The difference $ k= q_1-q_2 $ serves as the
gluon momentum. There are two discontinuities $ \pm k $ of the local momentum
$p(s$ at the edge of the disk, therefore the operator $ L(p) = 2 k^2 $ when
applied to product $W_0[p_{12}] W_0[p_{21}] $ on the right side.

There are also some terms coming from the commutators of the functional
derivatives $ \ff{p(t_{1,2})} $ with $L(p)$. These terms drop in this order
after integration over the loop. So, when inverted operator $\inv{L(p)} $
commutes with these functional derivatives and reaches the product $W_0[p_{12}]
W_0[p_{21}] $, it yields the gluon propagator $ \inv{2 k^2} $.

In the third order we can no longer neglect the commutator terms. These terms
are equivalent to the $ [A_{\mu},A_{\nu}] $ terms in a field strength. They
produce the triple vertex in the planar graphs. It is amazing, that all these
planar graphs with all their ghost and gauge fixing terms are generated by such
a simple mechanism.

Each term is very singular, but as we argued in\ct{Mig86}, these are spurious
singularities, arising due to the nonperturbative origin of quark
confinement.  In fact, in a massive theory, there is no reason to expect
singularities at zero momenta.  The expansion coefficients in front of the
momenta grow as $ \EXP{c/g_0^2} $ which is why there are singularities in the
perturbation expansion.

If we assume that there is an expansion in powers of momenta, the natural mass
scale $ M$ would emerge from the zero order  term $W(0) = M^4 $.
The precise meaning of this expansion is established in the next section. We
could use the scale invariance of the momentum loop  equation to express the
remaining terms in powers of $W(0)$.

\subsection{Concluding Remarks}

Another point to clarify. On the right hand side one should differentiate both
$ W$ functionals, according to the identities\footnote{Another way to arrive at
the same point splitting prescription is to follow the Feynman's prescription
of placing $A_{\mu}(x) $ in the middle of the interval $ dx$ in the loop
product.  Then $ dx = x(t+\oh dt) - x(t-\oh dt) $ which leads to the half of
the sum of the left and right velocities $ dx = \oh x'(t+0) dt + \oh x'(t-0) dt
$.  The velocities $ x' $ then transform to the functional derivatives in $ p $
space.}
\be
\ff{p_{\mu}(t_i)} = \oh \ff{p_{\mu}(t_i+0)} + \oh\ff{p_{\mu}(t_i-0)}.
\ee
The resulting four terms reduce to two inequivalent ones
\be
2 W'W' + 2 W''W .
\ee

One could introduce the area derivative in momentum loop space
\be
\ff{p_{\mu}(t)} = p'_{\nu}(t) \ff{\sigma_{\mu\nu}(t)},
\ee
which would make the parametric invariance explicit.

So, finally, the momentum loop equation takes the form
\bea
W[p]\,\delta_{\alpha\beta}\, \int_0^1 d t_2 p'_{\beta}(t_2) \int_0^{t_2} d t_1
p'_{\alpha}(t_1) =
\int_0^1 d t_2 p'_{\beta}(t_2) \int_0^{t_2} d t_1 p'_{\alpha}(t_1) \br
\left[ \fbyf{W\left[p_{12}\right]}{\sigma_{\mu\alpha}(t_1+0)}
\fbyf{W\left[p_{21}\right]}{\sigma_{\mu\beta}(t_2+0)}
+ W[p_{12}] \frac{\delta^2 W\left[p_{21}\right]}{\delta
\sigma_{\mu\alpha}(t_1-0)\delta \sigma_{\mu\beta}(t_2+0)}\right].
\eea
We could depict it as follows

\input epsf \centerline{ \epsfbox{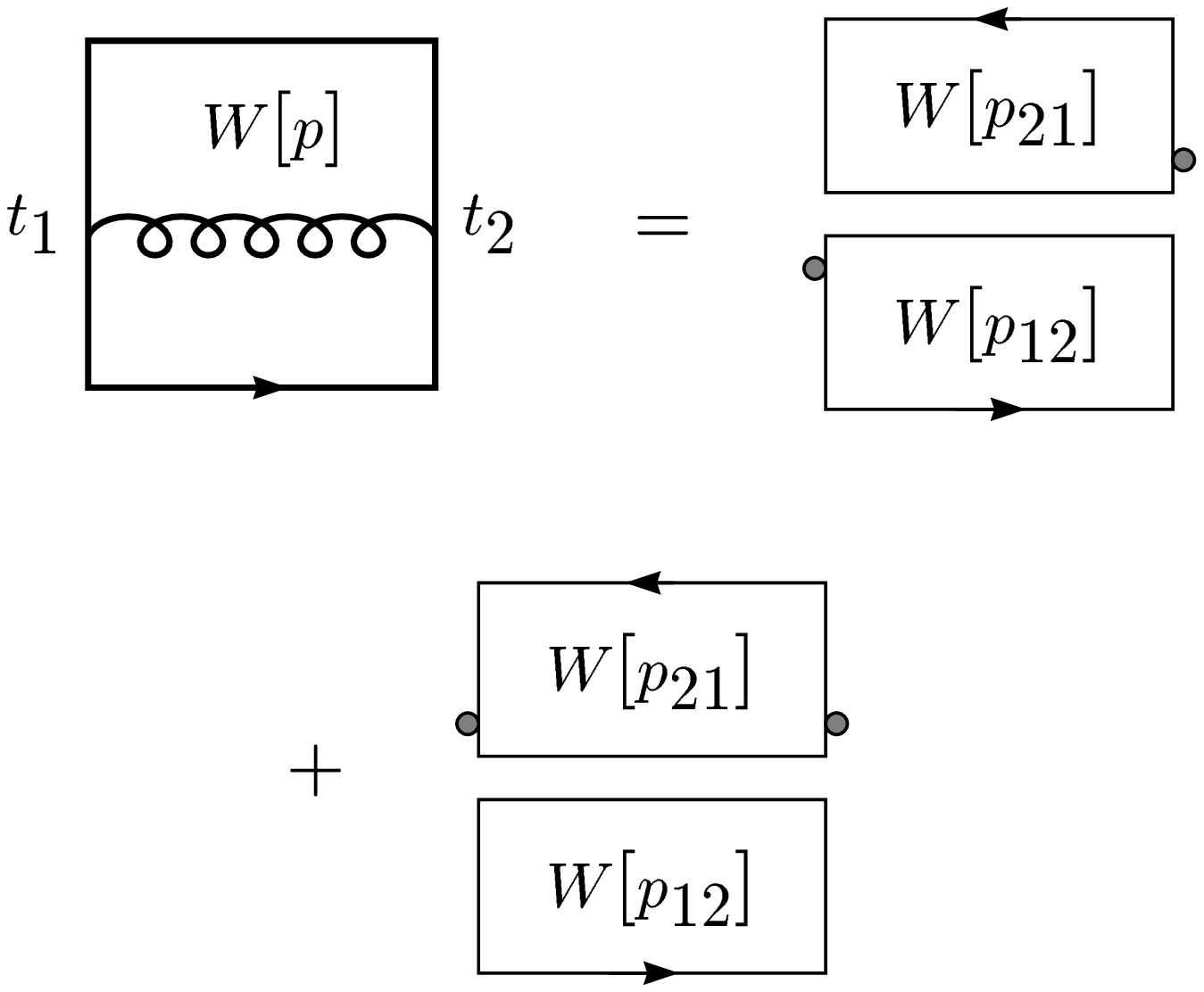}}

It is tempting to assume that the integrands on the both sides coincide, but
this is not necessarily true, as the perturbation theory shows. There are some
terms, which vanish only after the closed loop integration. In the one gluon
order these are the $ k_\mu k_\nu $ terms in the gluon propagator. In general,
these are the terms, related to the gauge fixing.

\section{Models of the Momentum Loop}

In this section we discuss various models and limits of the momentum loops to
get used to this unusual object. In particular, we consider the QCD2, where the
loops are classified by topology, and we also consider perturbative QCD in
arbitrary dimension by WKB approximation.

\subsection{``Exact'' Solution}

Let us first resolve the following paradox. There seems to be an "exact
solution" of this equation of the form
\be
W_?[p] =  Z \EXP{ \oh G_{\mu\nu} \oint p_\mu d p_\nu}.
\ee
Substituting this Anzatz into the \RHS we use the relations
\be
\fbyf{W_?[p]}{\sigma_{\mu\nu}(s)} =  G_{\mu\nu}\, W_?[p]
\ee
and note that the tensor area is additive
\be
\oint_P p_\mu d p_\nu = \int_{P_{12}}p_\mu d p_\nu \, +  \, \int_{P_{21}}p_\mu
d p_\nu,
\ee
where $ P_{12}$ and $P_{21}$ are parts of original loop $P$, closed with the
straight lines. These straight lines cancel in the tensor area integral, as
they
enter with opposite orientation.

The exponentials trivially reproduce the \LHS, as for the factors in front,
they yield
\be
2 Z^2 \int_0^1 d t_2 p'_\beta(t_2)\int_0^{t_2} d t_1 p'_\alpha(t_1)
G_{\mu\alpha}  G_{\mu\beta}.
\ee
This would agree with the \LHS provided
\be
2 Z G_{\mu\alpha}  G_{\mu\beta} = \delta_{\alpha\beta}.
\ee

The rotations reduce the antisymmetric tensor $G_{\mu\nu}$ to the canonical
form
\be
G_{2k-1,2k} = -G_{2k,2k-1} = \pm \inv{\sqrt{2 Z}}.
\ee
with the rest of elements vanishing.

Clearly, something went wrong. The formal Fourier transformation would bring us
back to the constant abelian field, which cannot solve QCD.

When we take a closer look at this Fourier transformation, we find out what
went wrong. The Gaussian integral reduces to the solution of the classical
equation
\be
G_{\mu\nu} p'_\nu(s) = \i x'_\nu(s),
\ee
which implies that $p(s)$ could have no discontinuities, since the loop $x(s)$
is continuous.

In order to reproduce the gluons we must satisfy the loop equations for the
discontinuous momentum loops as well. In gluon perturbation theory, the Fourier
integral involves the sum over number $n$  of discontinuities and integrals
over their locations $ s_i $
\bea
W[C] = \sum_{n=0}^\8 \int d s_1 \dots d s_n \theta\lrb{  s_1 < s_2 \dots < s_n}
\br
\int \lrb{DP}_{s_1,\dots, s_n} W_{s_1,\dots, s_n} [p] \EXP{\i \oint p_\mu(t) d
x_\mu(t)}.
\eea
These amplitudes $W_n[p] $ are related to the the gluon $n-$ point functions
$G_{\mu_1\dots\mu_n}\lrb{k_1,\dots k_n} $ as follows
\bea
W_{s_1,\dots, s_n} [p] = \br
G_{\mu_1\dots\mu_n}\lrb{k_1,\dots k_n} \delta\lrb{\sum k_i} \i^n
\ff{p_{\mu_1}(s_1)} \dots \ff{p_{\mu_n}(s_n)} \delta\left[p(\tau)-\sum k_i
\theta(\tau-s_i)\right].
\eea
Clearly, the sum here could diverge. We expect the nonperturbative $W[p]$ to be
a smooth {\em functional} of momenta (nevertheless these momenta could be
non-smooth {\em functions} of their proper time variables).

In the case of the constant abelian field, performing the Fourier
transformation from $x$ to $p $ space with $n$ discontinuities $ \Delta p(s_i)
=k_i$ we would get
\bea
\int D x \EXP{\oh \i F_{\mu\nu} \oint x_\mu d x_\nu - \i \oint x_\mu d p_\mu -
\i \sum x_\mu(s_i) k_i} \br
\propto \EXP{ - \oh \i (F^{-1})_{\mu\nu} \oint p_\mu d p_\nu} \prod_{i=1}^n
\delta^d(k_i).
\eea
The product of the delta functions comes from the integrations over the
positions of $ x_i $. This singularity arises because we have no $ \dot{x}^2 $
term in the exponential to damp spikes in $x(s) $.

So, the simple exponential in momentum space does not correspond to the
constant abelian field. It would rather correspond to violation of the gauge
invariance, as we could only obtain it by integrating over the discontinuous
loops in coordinate space. Therefore we have to reject this formal solution.

\subsection{Perturbative QCD}

Let us now discuss the perturbative solution. The formal perturbation expansion
for the momentum Wilson loop does not reflect its asymptotic behavior. To get a
better description of the asymptotic freedom in momentum loop space, we should
rather expand in the exponential of the path integral
\be
W[p] \propto \int D x \EXP{\log(W(x)) -\i \oint p'_\mu(s) x_\mu(s)},
\ee
near some (imaginary) saddle point $x(s)= C(s)$. It should satisfy the equation
\be
\i p'_\mu(s)  = C'_{\nu}(s)\fbyf{\log{W(C)}}{\sigma_{\mu\nu}(s)}.
\ee
In the first nontrivial order we get an equation of the following form
\be
\i p'_\mu(s) = Ng^2(C)\, C'_\nu(s) \oint \left( d C_\mu(t) \d_\nu - d C_\nu(t)
\d_\mu \right) D\lrb{ C(s)-C(t)}.
\ee
Here the running coupling constant $Ng^2(C) $ corresponds to the scale of $C$
and
\be
D(x) = \MINT{q} \frac{e^{\i q x}}{q^2} \propto (x^2)^{1-\oh d},
\ee
is the gluon propagator.
We could take this classical equation as a {\em parametrization} of $ p_C(s) $.
We observe that $ p_C \sim Ng^2(C)|C|^{3-d} $.  In four dimensions, for the
small loop $C$, the corresponding momenta are large. In two and three
dimensions, large coordinate loops correspond to large momentum loops.

As for the powerlike divergency of the contour integral at $t \ra s$ in three
and four dimensions, it could be regularized by the principle value
prescription $x^2 \ra x^2 + \epsilon $. For the smooth nonintersecting loop $C$
such integrals are finite.

The $ \i \oint C_\mu d p_C^{\mu}  $ term  reduces to $ -2$ times the same one
gluon exchange graph in virtue of the classical equation. So, we get the
following large $p$ asymptotics for $W[p]$ in the $C$-parametrization
\be
\log W[p_C] =  + \oh Ng^2(C) \oint d C_\mu(s) \oint d C_\mu(t) D\lrb{C(s)-C(t)}
\sim  Ng^2(C) |C|^{4-d}.
\ee
Unlike the formal perturbation theory, this expression is a well defined
parametric invariant functional of $C(s)$. One could go further and compute the
quadratic form for the second variations $ \delta C(s) $ near the saddle point.
This would give the pre-exponential factor in $W[P_C]$.

Note that in two dimensions there is no pre-exponential part for the
non-intersecting loop. The first order of perturbation theory is exact for $
\log W(C) $ in this case. The integrals  can be done, and they reduce to the
(absolute value of) the tensor area. Thus,
\be
p_C^{\mu}(s) = \i\, Ng^2\, \epsilon_{\mu\nu}\, C^{\nu}(s),
\ee
and
\be
\log W[P] = -\inv{2 N g^2} \,\left|\oint p_1 d p_2 \right|,
\ee
up to the nonlocal corrections due to self-intersecting $C$-loops in the path
integral.

At present we do not know how to compute these corrections,  but at large $p_C$
the fluctuations $\delta C \sim \inv{\sqrt{Ng^2}} $ are small compared to the
classical loop $ |C| \sim \frac{|p_C|}{Ng^2} $. Therefore, for large $p$ we
have the area law for $W[p]$. This is the general phenomenon, as one can see
from the momentum loop equation. In fact, the above "derivation" of the tensor
area law is valid for the large loops, where discontinuities and
self-intersections are irrelevant.

\subsection{Beyond Perturbation Theory: Random Vacuum Field}

To get a rough idea what could be the  momentum Wilson loop beyond perturbation
theory, let us consider a simple "model"  of QCD vacuum as random constant
abelian field $F_{\mu\nu}$ with some probability distribution $P(F)$. As we
discussed above, this is not a solution of the loop equations. Such model
misses the hard gluon phenomena, responsible for asymptotic freedom, but it is
the simplest model with confinement.

The Wilson loop would be
\be
W(C) \sim \int P(F) dF \EXP{ \oh \i F_{\mu\nu} \, \oint x_\mu \, d x_\nu},
\ee
and the corresponding momentum loop (for continuous $p(s)$)
\be
W(p) \sim \int P(F) dF \sqrt{\det F}\,\EXP{ -\oh \i F^{-1}_{\mu\nu} \, \oint
p_\mu \, d p_\nu}.
\ee
The pfaffian $\sqrt{\det F} $ comes as a result of computation of the
functional determinant by the $\zeta$ regularization in Appendix A.

Both loop functionals could be expanded in Taylor series in this simple model.
The Taylor expansion at small $X$ exists, because there are no gluons, just the
constant field. This is a wrong feature, to be modified. The Taylor expansion
at small $p$ exists, because the Wilson loop decreases fast enough in this
model. This feature we expect to be present in the full nonperturbative QCD.

Note, that the Fourier integrals here are not the usual integrals with the
Wiener measure, corresponding to Brownian motion. These are the parametric
invariant integrals, without any cutoff for higher harmonics of the field $x(s)
$ or $p(s)$. Such integrals
can be defined as limits of the usual integrals, with discrete Fourier
expansion at the parametric circle.
\be
x(t) = \sum_{_\8}^{\8} a_{n} e^{2 \pi \i n t}.
\ee
For the readers convenience, we consider these integrals in Appendix. We also
compute the divergent pre-exponential factors with the $\zeta$ regularization.

The important thing to know about these integrals is that they exist. One does
not have to break the parametric invariance by first adding the gauge fixing
terms and then compensating that by some Jacobians. One could simply compute
them with some invariant regularization, such as $\zeta$, or Pauli-Villars
regularization.

This also regularizes the wiggles, or backtracking. The Wilson loop does not
change when a little spike with zero area (backtracking) is added to the loop.
Such backtracking are hidden in the high harmonics $a_n $ of the position
field. The regularization of these wiggles leads to finite result in the
constant field model. We expect the same in full nonperturbative QCD.

In this case, the  position field could exist, unlike  the usual master field.
Our Anzatz \rf{ANZATZ} yields for the Taylor expansion terms
\bea
W_n(p) \propto \inv{n!N}\,\tr \hat{T}  \left(\oint_C d p_\mu(s)
\X{\mu}\right)^n=\br
\inv{N}\,\int\dots \int \theta\lrb{s_1<s_2\dots<s_n} d p_{\mu_1}(s_1) \dots d
p_{\mu_n}(s_n) \, \trb{\X{\mu_1}\dots \X{\mu_n}}.
\eea
These terms, in fact, reduce to traces of products of commutators, as there is
translational symmetry $ X_\mu \Ra X_\mu + 1* a_\mu$.

Comparing the above path integral for these terms with the second quantized
version we conclude that the string correlators
\be
\inv{n!}\int D x \,\delta^d(x(s_0))\, W(C) \,x_{\mu_1}(s_1) \dots
x_{\mu_n}(s_n)
\ee
do not depend of proper times $s_i$, as long as these times are ordered. In
general case these are combinations of the step functions, as they should be in
parametric invariant (i.e. 1D topological) theory. Integrating these terms by
parts with $  d p_{\mu_1}(s_1) \dots  d p_{\mu_n}(s_n) $  we would reproduce
the commutator terms coming from derivatives of the step function
$\theta\lrb{s_1<s_2\dots<s_n}$ in the ordered operator product representation.

These string correlations correspond to the Neumann boundary conditions. Here
we do not fix the edge $C$ of the "random surface", but rather let it fluctuate
freely. This fixes the corresponding local momentum: $ p_\mu=0 $ at the edge of
the surface.

Adding the $\oh \alpha \dot{x}^2 $ term at the edge would correspond to the
term $ \oh \inv{\alpha} p^2 $ in the Hamiltonian.  We do have such terms in our
physical Hamiltonian, of course, but for the purposes of computation of the
simplest correlators we could set $ \alpha \ra 0 $, which is equivalent to the
boundary condition $ p=0 $.

Our approach to the "QCD string" is similar to the matrix model approach to the
2D quantum gravity. We shall work directly in parametric invariant "sector"
rather than fixing the gauge and then imposing constraints to come back to the
physical sector.

\section{Equations for Dual Amplitudes}

The momentum loop equation is still ill defined. The meaning of the functional
derivatives $\ff{p(t_i)} $ at the breaking points $t_1,t_2$ needs to be
clarified. Also, the necessity to include momentum discontinuities  makes the
discrete Fourier expansion useless. Ideal thing would be some nonsingular
equation with functions of finite number of variables instead of functionals.

\subsection{Kinematics}

Let us  try to obtain the closed equations for the dual amplitudes
\be
A\lrb{k_1,\dots k_n} = \frac{1}{N}\, \tr V(k_1) \dots V(k_n),
\ee
where
\be
V(k) = \EXP{\i k_\mu X_\mu}
\ee
is the vertex operator, and momenta $k_i$ add up to zero.

These  traces are cyclic symmetric, of course. In addition there are relations
between amplitudes with various number of points
\bea
A\lrb{\dots,\xi k,\eta k,\dots} = A\lrb{\dots,(\xi +\eta) k,\dots}
\eea
which do not have analogies in the particle theory.
In particular, the two point function is trivial
\be
A(k,-k) = A() =1.
\ee
The first nontrivial one is the three point function $A(k,q-k,-q)$.
Given the generic $n$ point function, we automatically get all the previous
functions by setting some momenta to zero.

The momentum loop represents the polygon with the sides $ k_i $ and vertices
\be
p_i = \sum_{j=1}^i k_j,\; k_i = p_i - p_{i-1}, p_0 =0.
\ee
In parameter space the sides are the discontinuities, they take infinitesimal
time steps, $ s_i < t < s_i  + \epsilon$. This does not matter, in virtue of
parametric
invariance. The only thing which matters is the geometric shape of this
polygon.

These amplitudes corresponds to $W[p]$ in the limit
\be
p(s) \ra \sum_{i=1}^n k_i \theta(s-s_i),
\ee
when there is no smooth part of momentum. Such limit is quite natural from the
point of view of the string theory, but in perturbative QCD it would be
singular.

Existence of such limit implies convergence of the path integrals with $W(C) $,
which is violated in formal perturbation theory. In above constant field model
such limit exist. The symplectic form becomes the quadratic form in momenta
\be
F^{-1}_{\mu\nu}\oint p_\mu d p_\nu  \Ra \sum_{i=1}^n F^{-1}_{\mu\nu}
k_i^{\nu}\sum_{j=1}^{i-1} k_j^\mu
\ee
Given the \PDF for the tensor $F$ one could compute the corresponding
amplitudes in this model.

\subsection{Loop Equations}

Let us turn to the loop equations in real QCD. We already know the limit of the
\LHS of the momentum loop equation. It is just
\be
L(W)(p) \ra \lrb{\sum_{i=1}^n k_i^2} \,A\lrb{k_1,\dots k_n}.
\ee
This involves the sum of squares of lengths of the polygon.

The problem it to find the limit of the \RHS of the loop equation. We have to
compute it for the general Anzatz\rf{ANZATZ}, and then tend the local momentum
to the sum of the theta functions.

The double integral around our polygon breaks two of its sides

\input epsf \centerline{ \epsfbox{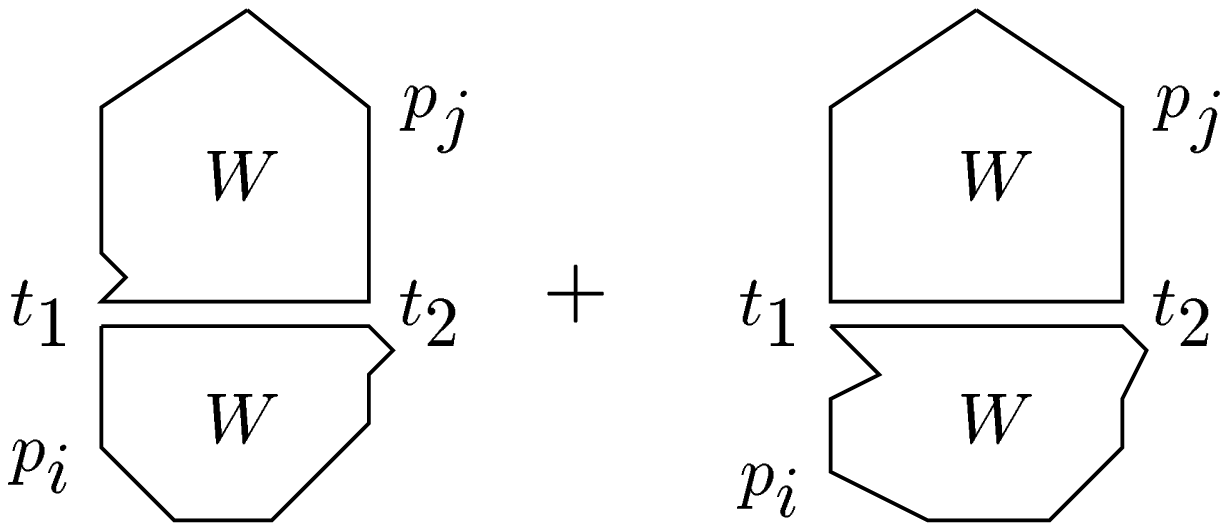}}

The extra discontinuity of momentum
\be
\Delta = p(t_1)-p(t_2)
\ee
arising with opposite signs in the two momentum loops leads to extra vertex
operators
\bea
W[p_{12}] = \inv{N}\,\trb{\U{t_1}{t_2}V(\Delta)},\br
W[p_{21}] = \inv{N}\,\trb{\U{t_2}{t_1}V(-\Delta)}.
\eea
These extra vertex operators correspond to internal sides, closing these two
loops. They are not parts of external loop $ p $, so they are not involved in
the functional derivatives.

Let us now apply the functional derivative $\ff{p(.)}$ to our Anzatz. Start
with the identity
\be
-\i\ff{p'_\nu(u)} \U{s}{t} = \U{s}{u} \X{\nu} \U{u}{t},
\ee
and differentiate it in $u$. We find
\bea
-\i\ff{p_\nu(u)} \U{s}{t} = \i \pp{u}\ff{p'_\nu(u)} \U{s}{t} \br
= p'_\alpha(u) \U{s}{u} [\X{\nu},\X{\alpha}] \U{u}{t}.
\eea
These are momentum space analogs of the Mandelstam relation for the area
derivative, with $X_\mu $ playing the role of the gauge field.
These identities play the crucial role in our construction. Without them we
would have only functional equations, for the general function $p(s)$.

When the momentum tends to the sum of the theta functions, the integrals over
$t_1,t_2$ are concentrated inside the infinitesimal intervals $ (s_i, s_i+
\epsilon) , (s_j, s_j + \epsilon)$ , where derivative $ p'$ is present. Due to
our principle value prescription they can never fall in the same interval.
Inside such intervals the functional derivatives split the vertex operator
\be
\delta V(k_i) \propto \int_{s_i}^{s_i+\epsilon} d p_\alpha(t)
\lrb{V(k'_i)[\X{\nu},\X{\alpha}] \otimes
V(k''_i)+V(k'_i)\otimes[\X{\nu},\X{\alpha}] V(k''_i)} .
\ee
where
\be
k'_i = p(t)-p(s_i), k''_i= p(s_i+\epsilon) -p(t), k'_i+k''_i=k_i.
\ee
The first factor in the tensor product goes in one trace, the second one goes
in another trace, since the loops reconnect precisely at the variation point $t
=t_1$ or $t=t_2$. The sum of  two terms with commutators switching from one
trace  to another arises because  our definition of velocity $x'(u)$ involves
the sum of left and right derivatives.

Inside each infinitesimal interval $ dp(t) = k d\xi, k'= \xi k,k''=(1-\xi)k$,
where $\xi = \frac{t-s_i}{\epsilon} $.  Therefore we get
\be
\delta V(k_i) \propto k_i^\alpha\int_0^{1} d \xi \lrb{V(\xi
k_i)[\X{\nu},\X{\alpha}] \otimes V((1-\xi)k_i)+V(\xi
k_i)\otimes[\X{\nu},\X{\alpha}] V((1-\xi)k_i)} .
\ee
The new objects here are the commutators. However, they can be obtained as a
limit of the vertex operators
\be
[\X{\mu},\X{\alpha}] \, k_{\alpha} =  2\lrb{\frac{\d^2}{\d{q_\mu}\d{\rho}}
V(\rho k)V(q-\rho k)V(-q)}_{\rho=q=0}
\ee
Geometrically this corresponds to adding little triangle in the $\mu $
direction as shown at the above figure.

Now we have all the ingredients in place. We obtain the following equation
\bea
\label{CHAIN}
\lrb{\sum_{l=1}^n k_l^2}\, A(k_1,\dots k_n) +
4\, \sum_{i\neq j}
\int_0^1 d \xi \,\int_0^1 d \eta \,
\lrb{\frac{\d^4}{\d q_1^\mu\d q_2^\mu \d\rho_1\d \rho_2}}_{q_i=\rho_i=0} \br
A\lrb{\eta k_j, k_{j+1}, \dots,(1-\xi+ \rho_1)k_i,q_1- \rho_1
k_i,-q_1,\Delta}\br
A\lrb{\xi k_i, k_{i+1}, \dots ,(1-\eta+ \rho_2)k_j,q_2- \rho_2
k_j,-q_2,-\Delta}
+\br
A\lrb{-q_1,q_1- \rho_1 k_i,(\xi + \rho_1) k_i, k_{i+1}, \dots ,(1-\eta+
\rho_2)k_j,q_2- \rho_2 k_j,-q_2,-\Delta }\br
A\lrb{\eta k_j, k_{j+1}, \dots ,(1-\xi)k_i,\Delta}\br
=0.
\eea
where
\be
\Delta = \xi k_i + \dots + (1-\eta)k_j = -\lrb{\eta k_j +\dots +(1-\xi)k_i}
\ee
is the transfer momentum.

These are exact equations for functions of finite number of variables rather
than functionals. The main advantage of this equation over the old loop
equation is absence of singularities. There are no delta functions here, the
only $k$-dependence is polynomial.

\section{Taylor Expansion at Small Momenta}

One may  expand these amplitudes in momenta (which should be possible in
massive theory), and recurrently find the expansion coefficients.
An important property of these equations  is the fact that the same
coefficients   serve  {\em all } the amplitudes, for arbitrary $n$.

\subsection{Equations for Commutators}

These expansion coefficients are related to the traces of products of the
$\X{}$ operators. Such traces are universal tensors, composed  from
$\delta_{\mu\nu} $, independently of the number of $k_i$ variables multiplying
these tensors.

We shall denote by $A_L$ the $\O{k^L}$ terms in the amplitude, for generic
momenta $ k_1,\dots k_n$. In fact, this $A_L$ would involve only combinations
of $L $ or less different momenta, so that remaining momenta are redundant.
This allows us to find relations for these coefficients, taking only finite
number of momenta.

The leading terms are quadratic in momenta.  Matching these two quadratic forms
gives the first nontrivial  relation. The $A$ term yields $ \sum k_l^2$. In the
remaining $ A \otimes A$ terms only the product of the zeroth  and fourth order
terms $A_4$ contribute. The fourth order terms reduce to the trace of two
commutators
\be
\frac{1}{N} \tr [\X{\mu},\X{\alpha}]\,[\X{\nu},\X{\beta}] =
a\, \lrb{\delta_{\mu\nu}\delta_{\alpha\beta}-
\delta_{\mu\beta}\delta_{\nu\alpha}}.
\ee
We find the following sum rule\footnote{It was first obtained by different
method in collaboration with Ivan Kostov(unpublished).}
\be
2(d-1)a  = 1
\ee

In higher orders we get more relations, each time involving the higher
derivative terms, i.e. higher commutators of position operator.
In order to study these relations systematically, let us parametrize the
conserved momenta as $ k_i = p_{i} - p_{i-1} $ and rewrite the dual amplitude
as follows
\be
A\lrb{p_1-p_n,\dots,p_n-p_{n-1}} = \inv{N} \trb{S(p_1,p_n)\dots S(p_n,p_{n-1})}
\ee
with
\be
S(k,q) = V(q) V(k-q) V(-k) = \TEXP{\i \int_{\triangle}X_\mu d p_\mu}
\ee
where the ordered integral goes around the triangle made from three vectors
$q,k-q,-k$. These triangles fill the polygon with vertices $p_i $ and sides
$k_i $.

At small momenta this formula reduces to
\be
S(k,q) \ra 1 + \oh k_\mu q_\nu \,[\X{\mu},\X{\nu}] +
\frac{\i}{6}\lrb{k_\mu k_\nu q_\lambda - q_\mu q_\nu
k_\lambda}[\X{\mu},[\X{\nu},\X{\lambda}]] + \dots.
\ee
This is the nonabelian version of the Stokes theorem for the triangle. One can
find higher terms here using the ordered integrals. From this representation it
is obvious that the Taylor expansion starts from the quartic term, as the trace
of commutator vanishes.

In higher orders we get  the  traces of polynomials to the set of traces of all
possible commutators
\be
\X{\nu_1\nu_2\dots\nu_n} \equiv [\X{\nu_1},[\X{\nu_2},\dots \X{\nu_n}]\dots].
\ee
The general translation invariant trace would involve products
\be
T\lrb{\{\nu\}_n,\{\mu\}_m \dots} =  \tr \X{\nu_1\nu_2\dots\nu_n}
\X{\mu_1\mu_2\dots\mu_m} \dots.
\ee

These $T$-traces  are not independent tensors. Those would be given by the
usual traces
\be
G\lrb{\mu_1\dots\mu_n} = \inv{N}\,\tr \X{\mu_1}\dots \X{\mu_n},
\ee
except these traces do not exist because they are not invariant under
translations $\delta \X{} = a *1 $.

\subsection{Planar Connected Moments}

The following parametrization proved to be convenient. Let us introduce the
{\em planar connected} moments $ W_{\mu_1\dots\mu_n} $ defined recurrently as
follows
\bea
W_\mu = G(\mu); \br
W_{\mu_1\dots\mu_n} = G\lrb{\mu_1\dots\mu_n}
- \sum_{planar} W_{\mu^{(1)}} \dots W_{\mu^{(m)}}.
\eea
Here the sum goes over all planar decompositions of the set $
\lrb{\mu_1\dots\mu_n} $ into subsets $ \lrb{\mu^{(1)}}, \dots \lrb{\mu^{(m)}}
$. For example,
\be
W_{\mu  \nu}= G(\mu \nu) - G(\mu) G(\nu),
\ee
\be
W_{\mu  \nu \lambda} = G(\mu \nu \lambda)
- G(\mu) G(\nu \lambda) - G(\nu) G(\lambda \mu) - G(\lambda) G(\mu\nu)
+ 2 G(\mu) G(\nu) G(\lambda).
\ee
In the next section we shall  use this representation to build the Fock space
representation of the $X$ operator.

The advantage of these $C-$ moments is their simple behaviour under
translations. Only the first one gets translated
\be
\delta \,W_\mu= a_\mu
\ee
while all the higher ones stay unchanged.
This can be proven by induction. The variation of the $G-$ moments under
infinitesimal translation reads
\be
\delta \, G\lrb{\mu_1\dots\mu_n} = \sum_l a_{\mu_l}
G\lrb{\mu_{l+1}\dots\mu_{l-1}},
\ee
with the cyclic ordering implied.
Straightforward planar graph counting gives the same for the variation of the
planar representation
\bea
G\lrb{\mu_1\dots\mu_n} = W_{\mu_1\dots\mu_n}
+ \sum_{planar} W_{\mu^{(1)}} \dots W_{\mu^{(m)}}, \br
\delta\, G\lrb{\mu_1\dots\mu_n} = \sum_{planar}
W_{\mu^{(1)}} \dots \delta W_{\mu_l} W_{\mu^{(m)}} = \br
 \sum_l a_{\mu_l}  \sum_{planar}  W_{\mu^{(1)}} \dots W_{\mu^{(m)}} = \br
\sum_l a_{\mu_l} G\lrb{\mu_{l+1}\dots\mu_{l-1}}
\eea
In the next section we present a simpler proof based on operator
representation.

So, we could fix the gauge by imposing
\be
W_\mu= 0
\ee
and parametrize the traces by the invariant coefficients in front of various
terms in the planar connected moments. Those are constant tensors, composed
from $ \delta_{\mu_i\mu_j} $.
To simplify the formulas we shall use the notation
\be
\delta_{i,j} \equiv \delta_{\mu_i,\mu_j}.
\ee
This shall never lead to confusion, as all our indexes would be summed over at
the end anyway.

Only cyclic symmetric tensors may appear. Up to the eighth order
\bea
W_{1,2} = \delta _{1,2}\,A, \br
W_{1,2,3,4}=
\delta _{1,3}\,\delta _{2,4}\,B +
  \left( \delta _{1,4}\,\delta _{2,3} +
     \delta _{1,2}\,\delta _{3,4} \right) \,C ,\br
W_{1,2,3,4,5,6} =
\delta _{1,4}\,\delta _{2,5}\,\delta _{3,6}\,D + \br
  \left( \delta _{1,4}\,\delta _{2,6}\,\delta _{3,5} +
     \delta _{1,5}\,\delta _{2,4}\,\delta _{3,6} +
     \delta _{1,3}\,\delta _{2,5}\,\delta _{4,6} \right) \,E +\br
  \left( \delta _{1,6}\,\delta _{2,5}\,\delta _{3,4} +
     \delta _{1,2}\,\delta _{3,6}\,\delta _{4,5} +
     \delta _{1,4}\,\delta _{2,3}\,\delta _{5,6} \right) \,F + \br
  \left( \delta _{1,5}\delta _{2,6}\delta _{3,4} +
     \delta _{1,6}\delta _{2,4}\delta _{3,5} +
     \delta _{1,3}\delta _{2,6}\delta _{4,5} +
     \delta _{1,5}\delta _{2,3}\delta _{4,6} +
     \delta _{1,2}\delta _{3,5}\delta _{4,6} +
     \delta _{1,3}\delta _{2,4}\delta _{5,6} \right) \,G + \br
  \left( \delta _{1,6}\,\delta _{2,3}\,\delta _{4,5} +
     \delta _{1,2}\,\delta _{3,4}\,\delta _{5,6} \right) \,H.
\eea

{}From the above solution for the fourth moment we find
\be
B = C + A^2 + \inv{2 (d-1)}.
\label{U4}
\ee
The coefficients $A, C $  remain arbitrary.

\subsection{Higher Orders}

In higher orders, let us pick out the highest $\O{k^L}$ order term in
\rf{CHAIN}, moving the lower ones to the right
\bea
\sum_{i\neq j} \,k^{\alpha}_i\,k^{\beta}_j\,
\int_0^1 d \xi \,\int_0^1 d \eta \, \br
\trb{[\X{\nu},\X{\alpha}] V(\Delta)[\X{\nu},\X{\beta}]V(\eta k_j)V(k_{j+1})
\dots V((1-\xi)k_i)}_{k^{L-2}}\br
= F\lrb{A_{l<L-2},k_j}.
\eea
The left side involves (linearly!) the highest $C$ coefficients, with
$L+2$ indexes. The order of the coefficients on the right side is $ L
$ and lower.

So, this chain of equations have triangular form, which is a great
simplification. The number of equations we get in the $\O{k^L}$ order is equal
to the number of the  $L$-th order cyclic symmetric invariants made
from $n$ momenta.  The number of unknowns  coincides with the number
of cyclic symmetric tensors with $L+2 $indexes, made from
$\delta_{\mu\nu}$.

There are, in general, much more equations than unknowns, but some
hidden symmetry which reduces this system to {\em just one equation},
at least up to the orders we computed.

Thus, all the  $\O{k^2}$ equations for any number $n$ of legs reduce to
\rf{U4}.
The $\O{k^4}$ equations we studied, with up to $6$ legs, all reduce to the
following equation
\be
3 \,D =
{2\,{A}^3} + A\,\left( \frac{7}{2(d-1)} - 4 C \right)  + E - 4\,F +
7\,G - H.
\label{U6}
\ee
These equations were obtained by means of \MAT. We generated all cyclic
symmetric tensor structures recurrently, then introduced coefficient

in front and substituted into the operator expansion. The Cuntz algebra was
implemented to compute the expectation values of the products of the $\hat{\cal
X} $  operators.

We had to use some \MAT tricks to increase efficiency in computations of the
loop equations. In its present form the program could compute  a couple more
orders without much effort at the personal workstation. The program is
included in the uuencoded hep-th files and could also be provided by e-mail
upon  request.

In the next order we found the equation, but it is too lengthy to
write it down here. It is included in the \MAT file.

We could not find the general formula for the number of parameters in
$d$ dimensions. The group theory must have these answers.
Clearly, one cannot alternate more than $d$ indexes, which restricts invariants
made from $d+1$ or more $\delta_{\mu\nu} $ tensors.

In four dimensions this would start from ten indexes which we do not consider
here.  However, in two dimensions these restrictions are quite important.

\subsection{Symplectic Symmetry in QCD2}

There is, in fact more symmetry in two dimensions.
The symplectic invariance of 2D QCD corresponds to the following
transformations
\bea
\delta x_\mu(t) =  \alpha \epsilon_{\mu\nu} x_\nu(t),\br
\delta p_\mu(t) = -\alpha \epsilon_{\mu\nu} p_\nu(t)
\eea
One can readily check that these transformations preserve the symplectic form
$$
\oint p_\mu d x_\mu ,
$$
as well as the tensor areas
$$
\oint \epsilon_{\mu\nu} x_\mu d x_\nu,
\oint \epsilon_{\mu\nu} p_\mu d p_\nu.
$$
We already saw this symmetry in the WKB approximation at large $p$ when the
momentum loop was a function of the tensor area in momentum space.

These transformations lead to the symmetry of our equations in 2D
\be
\delta \X{\mu} = \alpha \epsilon_{\mu\nu} \X{\nu}.
\ee
As a consequence of this symmetry, one should use only the $\epsilon_{\mu\nu}$
symbols in constructing the expectation values of  products of commutators of
$\X{}$. The parity implies that the $\epsilon$ symbols come in pairs, therefore
only the multiples of four $\X{}$ operators could have nonzero traces.
The lowest commutator $\X{\mu\nu}= [\X{\mu},\X{\nu}] $ is symplectic invariant,
but the higher ones, in general, are not. This leads to extra restrictions on
the invariant structures, to be used in the solution.

Up to the eighth order the symplectic invariant solution in two dimensions
reads
\be
A = 0,\;B = \inv{3},\;C = -\inv{6} \dots
\ee
It corresponds to
\be
W_{\alpha\beta\mu\nu} = \inv{6} \,\lrb{\epsilon_{\alpha\beta} \epsilon_{\mu\nu}
+ \epsilon_{\beta\mu} \epsilon_{\nu\alpha}}.
\ee

\subsection{Higher Dimensions}

In higher dimensions it is tempting to try something similar. Due to the $Z_4$
symmetry
\be
p_\mu(s) \Ra \omega p_\mu(s), \;\omega^4 = 1,
\ee
of the  momentum loop equation there is always a solution with
$W_{1,\dots,n} = 0$ unless $ n = 0 mod 4 $, so that the Taylor expansion goes
in $p^4$.

This is not sufficient to eliminate the higher invariants in $C-$ tensors. Say,
in the fourth order the second invariant $C $ would still be present
\be
B = C  + \inv{2 (d-1)}.
\ee
Similar terms would exist in higher orders. We do not have symplectic symmetry
here to reduce invariants, so what do we do?

In principle, we should fix the unknown coefficients by matching with
perturbative QCD at large momenta. This is easier said than done.
As a model, we could try to build the general terms from commutators
$\X{\mu\nu} $ in arbitrary dimension.
Say, in the fourth order this would correspond to
\be
C = -\oh B,
\ee
so that the $C-$ tensor would be equivalent to the double commutator
\be
W_{1,2,3,4} = \oh \,B \lrb{ 2\,\delta _{1,3}\,\delta _{2,4} - \delta
_{1,4}\,\delta _{2,3} - \delta _{1,2}\,\delta _{3,4} }.
\ee
Solving the loop equation we find
\be
B =  \inv{3 (d-1)}.
\ee

In four dimensions we could get some number of Taylor terms analytically. Then,
the Pade approximation could be used to estimate the spectrum. One could also
try to compare some string models with these relations.

In Appendix B we discuss the possible numerical solution of these equations.
We postpone the numerics till the better understanding of analytic properties
of the momentum loop equation.

\section{Operator Representation}

The equations of the previous section can be used to recurrently determine the
planar connected moments $ W_{\mu_1\dots\mu_n}$.

Suppose we solved this algebraic problem. The natural next question is how to
reconstruct the matrix $\X{}$ from its moments. In the large $ N$ limit this
problem can be explicitly solved by means of the noncommutative probability
theory.

\subsection{Noncommutative Probability Theory}

The noncommutative probability theory in its initial form\ct{Voiculesku}
covered only so called free noncommutative variables, corresponding to the
factorized measure. The  theory of interacting noncommutative variables which
we use here, was developed in the recent work\ct{Gross94}, where  it was
applied to the matrix models and QCD2. Some elements of this theory were
discovered back in the eighties \ct{Cvitanovic} in the context of the planar
graph \SD equations.

Let us introduce auxiliary operators $ a_\mu $ and $ a^{\dagger}_\mu $
satisfying the  Cuntz algebra
\be
 a_\mu a^{\dagger}_\nu = \delta_{\mu\nu}
\ee
which is a limit of the $q-$ deformed oscillator algebra
\be
a_\mu a^{\dagger}_\nu  - q \, a^{\dagger}_\nu a_\mu = \delta_{\mu\nu}
\ee
when $ q \ra 0 $.
These operators correspond to the source $j = a^{\dagger} $ and noncommutative
derivative $ \ff{j} = a $ introduced by Cvitanovic et. al.\ct{Cvitanovic}.

We could explicitly build these operators as the limit of the large $N$ matrix
$ a^{\dagger}_\mu$ and its matrix derivative $ a_\mu = \inv{N}
\pp{a^{\dagger}_\mu} $. The trick is that we shall keep all the traces of
products of $a^{\dagger} $  {\em finite} in the large $N$ limit, so that the
moments vanish except the zeroth
\be
 \inv{N} \tr a^{\dagger}_{\mu_1} \dots a^{\dagger}_{\mu_n}  = \delta_{n0}.
\ee

The matrix derivative is the $N=\8 $ counterpart of the matrix of the
derivatives with respect to the matrix elements of $a_\mu$
\be
\lrb{a_\mu}^{ij}  = \inv{N} \pp{\lrb{a^{\dagger}_{\mu}}^{ji}}.
\ee
The transposition of indexes $ ij$ is needed to obtain the simple matrix
multiplication rules
\bea
\left(a_\mu a^{\dagger}_{\mu_1} \dots a^{\dagger}_{\mu_n}\right)_{ik}  =
\lrb{a_\mu}^{ij} \lrb{a^{\dagger}_{\mu_1} \dots a^{\dagger}_{\mu_n}}_{jk} = \br
\delta_{\mu\mu_1} \lrb{a^{\dagger}_{\mu_2} \dots a^{\dagger}_{\mu_n}}_{ik} +
\inv{N}
\sum_{l=2}^n \delta_{\mu\mu_l}\,\tr \lrb{a^{\dagger}_{\mu_1} \dots
a^{\dagger}_{\mu_{l-1}}} \lrb{a^{\dagger}_{\mu_{l+1}} \dots
a^{\dagger}_{\mu_n}}_{ik} \ra \delta_{\mu\mu_1} \lrb{a^{\dagger}_{\mu_2} \dots
a^{\dagger}_{\mu_n}}_{ik}
\eea
In the first term the factor of $N$ in definition of the matrix derivative
cancelled with the trace $ \delta_{ii} = N$. The remaining terms all contain
the vanishing moments.

\subsection{Operator Expansion for the Position Operator}

Let us define the following {\em operator expansion}
\be
\hat{\cal{X}}_{\mu}  = W_\mu \,* \,1 + a_{\mu}  + \sum
W_{\mu\nu_1\dots\nu_n} a^{\dagger}_{\nu_n} \dots
	a^{\dagger}_{\nu_1}
\ee
where the sum goes over all  planar connected  moments $ C $ with all tensor
indexes.
The first term is a c-number. It takes care of the possible vacuum average of
the position operator. It drops in the dual amplitudes, as it commutes with the
rest of terms and corresponding exponent can be factored out of the vertex
operator
\be
\hat{\cal{V}}(k) = \EXP{\i k_\mu  \hat{\cal{X}}_{\mu}}
\ra \EXP{\i k_\mu W_\mu} \hat{\cal{V}}(k).
\ee
These phase factors all cancel due to momentum conservation, which allows us to
drop the $W_\mu-$ term from the position operator.

With this representation it is obvious that the planar connected moments
$W_{\dots} $ are all translational invariant, except for the first one, $
W_\mu$. The translation of the c-number term provides the correct shift of the
position operator, and there is no way the variation of higher terms could have
compensated each other, as they enter in front of different products of the $
a_\mu^{\dagger}  $ operators.

Consider the expectation values of the products of such operators with respect
to the "ground state" $ \left| 0 \right\rangle $ annihilated by $a_\mu$
\be
a_{\mu} \left| 0 \right\rangle = 0, \; \left\langle 0 \right| a^{\dagger}_\mu =
0.
\ee
With our matrix derivative representation of these operators
\be
\left| 0 \right\rangle = 1, \; \left\langle 0 \right| = \delta(a^{\dagger});
\ee
so that
\be
\left\langle 0 \right| a_{\mu_1} \dots a_{\mu_n} F(a^{\dagger}) \left| 0
\right\rangle  = \inv{N}\pp{a^{\dagger}_{\mu_1}} \dots
\inv{N}\pp{a^{\dagger}_{\mu_n}} F(a^{\dagger} \ra 0)
\ee

Clearly, only the terms with all $ a $ and $ a^{\dagger} $ mutually cancelled
will remain. This gives the usual decomposition of the moments into planar
connected ones
\be
\left\langle 0 \right|  \hat{\cal{X}}_{\mu_1} \dots \hat{\cal{X}}_{\mu_1}
\left| 0 \right\rangle = W_{\mu_1\dots\mu_n} + \sum W_{\dots} W_{\dots} + \dots
\ee
The details of this beautiful construction for matrix models and 2d QCD can be
found in \ct{Gross94}.

In our case, the coefficients of the operator expansion are governed by the
loop equations. Up to the sixth order
\be
\hat{\cal{X}}_{\mu}  =  a_{\mu}  + A\, a^{\dagger}_{\mu}
+ B \, a^{\dagger}_{\nu} \,a^{\dagger}_{\mu} \,a^{\dagger}_{\nu}  +
C \,\lrb{ a^{\dagger}_{\mu} \,a^{\dagger}_{\mu} \,a^{\dagger}_{\nu} +
a^{\dagger}_{\nu} \,a^{\dagger}_{\mu} \,a^{\dagger}_{\mu}} + \dots
\ee
with the coefficients $ A,B,C,\dots$ satisfying above equations
\rf{U4},\rf{U6}.

The simplest solution, discussed in the previous section, reads,
\be
\hat{\cal{X}}_{\mu}  =  a_{\mu} + \inv{6(d-1)} \, \left[a^{\dagger}_{\nu}
\,\left[a^{\dagger}_{\mu} \,a^{\dagger}_{\nu}\right]\right]  +
\O{(a^{\dagger})^7}
\ee

What could be the use of this expansion? Perhaps, some truncation procedure
using some trial functions can be found. Or else, one can always use the
analytic methods, by extrapolation of expansion for the physical observables by
the Pad\'e technique. We do not know all the expansion parameters, they are to
be fixed by matching with perturbative QCD at large momenta.

\subsection{Perturbative vs Nonperturbative Fock Space}

It is instructive to compare this construction with that of \ct{Gross94}. We
have here (at arbitrary dimension of space) much smaller Fock space than the
one found by Gopakumar and Gross for QCD2. There, the operators $a, a^{\dagger}
$ had the momenta as their arguments, which made the Fock space similar to the
space of the free gluons (except for the Boltzmann statistics instead of the
Bose statistics of the usual free gluon theory).

The present construction has much less degrees of freedom, in fact, no
continuous degrees of freedom at all. We simply have $d$ pairs $a_\mu,
a_\mu^{\dagger} $ which is like the Fock space of a single Boltzmann
oscillator. The states are spanned by all $d-$ letter words
\be
\left | \mu_1 \mu_2\dots \mu_n \right\rangle =
a^{\dagger}_{\mu_1}a^{\dagger}_{\mu_2}\dots a^{\dagger}_{\mu_n}\left | 0
\right\rangle.
\ee
This tremendous reduction of degrees of freedom (very much like the one in
Eguchi-Kawai model, but in continuum theory) takes place due to implied
confinement.

It is due to confinement that we are able to expand amplitudes in local
momenta, so that only the Taylor coefficients need to be described. The Taylor
expansion is much simpler that the Fourier expansion, where  the basic set of
functions are labelled by continuous momenta.

Another simplification came from unbroken parametric invariance. The general
Taylor expansion of the loop functional  would involve unknown coefficient
functions, which would result in the loop operators $a_\mu(s) $, like those of
the conventional string theory. Expanding those in discrete Fourier expansion
on a unit circle we would get the usual infinite set of the string oscillators
$ a_{\mu, n}$.

So, why is our Fock space smaller that that of the string theory? This is
because we did not fix the parametric gauge either. Should we break the
parametric invariance by adding the conventional $ \dot{x}_\mu^2 $ term in the
exponential in the definition of the path integral, (i.e. use the Wiener
measure instead of the flat one we use here), we would have all these problems
back again.

However, as we learned from the recent success of  the matrix models, the
quantum geometry problems can be solved by purely algebraic and combinatorial
methods, without any breaking of parametric invariance. The (perturbative) Fock
space of the Liouville theory in conformal gauge involves continuum fields, but
we know that there is a different picture, with only discrete degrees of
freedom.

This picture is the matrix model quantized by means of the noncommutative
probability theory. The Fock space in this picture involves only one Cuntz pair
$a, a^{\dagger}$. The perturbative Fock space was too big as it turned out
after exact solution. After that, the Liouville theory Fock space was also
reduced, by methods of the topological field theory.

So, we expect the same thing in the large $N$ QCD. The nonperturbative Fock
space must be much smaller than the perturbative one due to the color
confinement. The operators we build here are discrete because we are working in
purported confining phase of QCD, where the motion is finite, so that most of
the perturbative gluon degrees of freedom are eliminated.

By the same reason we switched from the original gauge field (the Witten's
master field) to the conjugate coordinate operator. The master field was
unbounded even in confining phase due to the hard gluons. The coordinate field
is bounded in confining phase, which is infinitely simpler to describe in our
picture than the free motion.

In fact, without confinement, our methods would be completely pointless. The
description of the plane wave of unconfined gluon in terms of Taylor
coefficients would be very inefficient.

The proof of quark confinement within such approach cannot be achieved, all we
can say is that exact equations for the dual amplitudes allow for a systematic
expansion in local momenta, which can be interpreted in operator terms.

Note however, that these same equations, being iterated from perturbative
vacuum instead, generate all the planar graphs of perturbative QCD. The
equations being nonlinear, we cannot say whether our Taylor expansion describes
the same solution, which matches perturbative QCD.

\section{ Hadron Spectrum}

In this section we derive the wave equations for the mesons and glueballs.
These equations correspond to  a different physics: stringy states described by
a long timelike Wilson loop  in meson case, and small fluctuations of the
Wilson loop in the glueball case. Still, in both cases, the wave equations can
be explicitly written down in terms of the position operator.

\subsection{Mesons}

The equations of the previous section are supposed to fix $\X{}$ up to the
overall gauge transformation $ \X{\mu} \Ra S^{-1} \X{\mu} S$.  Let us assume we
found the solution for  $\X{}$, numerically or analytically. What can be said
about the particle spectrum? Let us study the meson sector first.

Consider the limit, when the loop becomes an infinite strip. This corresponds
to the propagation of the meson made from quark and antiquaqrk. There are two
position operators $\X{\mu},\bar{\X{\mu}}$, and the corresponding loop function
can be regarded as tensor product of the two ordered exponentials
\be
\Psi\lrb{T,\X{},\bar{\X{}}} \propto \int D p(.) \int D \bar{p}(.)  \TEXP{
\int_{-\8}^T d t \lrb{\i\,\X{\mu} \dbyd{p_\mu}{t} - H}\oplus
\lrb{\i\,\bar{\X{\mu}} \dbyd{\bar{p}_\mu}{t} -\bar{H}}}
\ee
with the above Dirac Hamiltonians \rf{HDIR}.

Let us shift the time $ T\Ra T + d T $ and try to get the Schr$\ddot{o}$dinger
equation (in imaginary time). Introducing the  vectors $ q_\mu =
p_\mu(T+dT)-p_\mu(T)$ we get the vertex operators
\be
\TEXP{\i \int_T^{T+dT}\X{\mu} d p_\mu(t)} \ra \EXP{\i q_\mu \X{\mu}} = V(q).
\ee
Now, expanding the rest of the factors in $dT$
\be
\EXP{-dT\lrb{m + \i \gamma_\mu p_\mu(T+ dT)}} = \EXP{- d T\lrb{\hat{H} + \i
\gamma_\mu q_\mu}} \ra 1 - d T\lrb{\hat{H} + \i \gamma_\mu q_\mu}
\ee
and comparing the linear terms, we get the following
\be
-\dot{\Psi}\lrb{T,\X{},\bar{\X{}}}= \lrb{\lrb{H + \gamma_\mu \hat{A}_\mu}
\oplus  \lrb{\bar{H} + \bar{\gamma}_\mu \bar{\hat{A}}_\mu}} \,
\Psi\lrb{T,\X{},\bar{\X{}}},
\ee
where
\be
\hat{A}_\mu = \i \MINT{q} V(q) q_\mu = \d_\mu \delta(\X{}).
\ee
We also get some terms with the quark mass corrections, which we absorb into
the bare mass $m$.

The $\hat{A}_\mu$ term is the net contribution of the "QCD string". This is the
natural generalization of the derivative operator in the Dirac equation.
Incidentally, this is what we get for the Witten's master field.

For a finite matrix $ \X{}$ this term would be a sum of the (gradients of the)
delta functions, whereas in the large $N$ limit it is related to the density of
eigenvalues at small $\X{}$. The singularities at small loops would correspond
to the divergent integrals at small eigenvalues.

There are $d$ different sets of eigenvalues $\kappa_\mu$ and the same number of
different eigenvectors $\hat{\Omega}_\mu$, as the matrices $\X{\mu}
=\hat{\Omega}_\mu \kappa_\mu \hat{\Omega}_\mu^{\dagger}$ do not commute.
Therefore, the angular variables, corresponding to the eigenvectors  of each
$\X{\mu}$ would not decouple here.

With the operator language described in the previous section, the gradient of
the delta function can be defined by its matrix elements with some trial
functions in the Fock space of the operators $ a_\mu^{\dagger} $.

Once we know  the master field, the above Schr$\ddot{o}$dinger equation is
quite explicit. The states with fixed total momentum
\be
P^{tot}  = P^{string} + p + \bar{p}
\ee
are selected by condition
\be
\i P^{string}_\mu \Psi\lrb{T,\X{},\bar{\X{}}} = \lrb{\trb{\D{\mu}} +
\trb{\pp{\bar{\X{\mu}}}}}\Psi\lrb{T,\X{},\bar{\X{}}},
\ee
corresponding to total translation. The mass spectrum is given by stationary
solutions, i.e.
\be
\lrb{\lrb{H + \gamma_\mu \hat{A}_\mu} \oplus  \lrb{\bar{H} + \bar{\gamma}_\mu
\bar{\hat{A}}_\mu}} \, \Psi\lrb{\X{},\bar{\X{}}} = 0.
\ee

The appropriate trial states are superpositions of the vertex operators
\bea
\Psi_{k_1,\dots k_n,\bar{k}_1,\dots \bar{k}_m}\lrb{\X{},\bar{\X{}}} =
V\lrb{k_1}\dots V\lrb{k_n}\otimes \bar{V}\lrb{\bar{k}_1}\dots
\bar{V}\lrb{\bar{k}_m},\br
 \sum k_i + \sum\bar{k}_j = P^{string}.
\eea
The above wave operator acts linearly on the space of such states. Note, that
we are not considering the dual amplitudes now. The quark degrees of freedom
are present in this hamiltonian. Still, the vertex operators can be used as the
basis for the wave equation.
The computational aspects are beyond our present scope.

\subsection{Glueballs}

In the glueball sector the situation is different. The glueball wave function $
G(C) $ can be obtained of as  variation of the Wilson loop $G(C) = \delta W(C)
$, in the same way, as the variations of the Higgs condensate describe the
physical excitations in the phase with "broken" gauge symmetry.

This relation between $ G(C) $ and $W(C)$ can be formally derived from the loop
equation for the generalized loop average
\be
W_Q(C) = \VEV{ Q\, \trb{\TEXP{\oint_C A_\mu d x_\mu} }}
\ee
where $Q$ is some invariant operator with the fixed total momentum,
\be
Q = \int d^d y \EXP{-\i k y} \trb{P\lrb{F_{\mu\nu},\nabla_\alpha
F_{\beta\gamma},\dots}}
\ee
One could think of this $Q$ as an infinitesimal loop, moved around with the
plane wave weight $ \EXP{-\i k y}$.

The functional $W_Q(C) $ is not translation invariant, it is rather the
momentum eigenstate
\be
\oint d s \ff{x_\mu(s)} W_Q(C) = \i k_\mu W_Q(C).
\ee
as it follows from the translation $ y \ra y + a $ in above integral.

The Levy operator still yields the \YM equation inside the trace. In virtue of
the \SD equations this leads to the usual loop splitting terms plus the contact
terms, which we ignore, as we are interested in the residue in the pole at the
glueball spectrum $ k^2 = - m^2$.
The loop equation for this residue $G_Q(C)$ reads
\be
\label{GEQ}
L(G_Q)(C) = R(G_Q,W)(C) + R(W,G_Q)(C)
\ee

We used here the factorization theorem, keeping in mind that for finite
momentum  $k$ the expectation value of $Q$ vanishes. Hence, the leading term in
the average product of the three invariant operators reads
\bea
\VEV{Q\, \W{xy} \W{yx}} \ra \br
\VEV{Q\, \W{xy}}\VEV{ \W{yx}} +\br
\VEV{Q\, \W{yx}}\VEV{ \W{xy}}
\eea

Now, the loop equation \rf{GEQ} would be satisfied for the {\em linear
variation}
\be
G_Q(C) = \delta W(C,\X{}) \equiv W\lrb{C,\X{}+ \delta\X{}} - W(C,\X{}),
\ee
along the nonlinear loop equation. In other words, if the nonlinear loop
equation is satisfied, when the position field $\X{}$ is shifted by $\delta
\X{}$, then the variation of $W(C,\X{}) $ satisfies the linearized loop
equation \rf{GEQ}.

Such perturbation corresponds to the variation of the vertex operator
\be
\delta V(k) = \int_0^1 d\lambda\,V(\lambda k) \,\i k_\mu\, \delta \X{\mu}
\,V((1-\lambda)k).
\ee
with $\delta \X{\mu} $ represented by the  operator expansion
\be
\delta \hat{\cal X}_\mu = \delta W_\mu + \sum \delta W_{\mu\mu_1\dots\mu_n}
a_{\mu_n}^{\dagger}\dots a_{\mu_1}^{\dagger},
\ee

The set of equations for the perturbed amplitudes
\be
\delta A(k_1,\dots k_n) = \sum_{l=1}^n \,\i k_l^\mu \,\int_0^1 d \lambda \,\tr
\delta \X{\mu} \,V((1-\lambda)k_l)\, V(k_{l+1}) \dots V(k_{l-1}) \, V(\lambda
k_l)
\ee
would now provide the linear relations for the coefficients $ \delta W_{\dots}
$.
This variation should be the momentum eigenstate, i.e.
\be
\delta A(k_1,\dots k_n) \propto \delta^d(k-\sum k_i)
\ee
after which we get linear homogeneous system of equations, involving $ k_\mu$
as a parameter. The solvability conditions should provide the spectral equation
for $Q$. In virtue of the space symmetry this would quantize the square
$k_\mu^2 = - M^2$.

\section{Conclusion}

This is as close as we could get to the QCD string. We built the model of $d$
large $N$ matrices $\X{\mu}$ representing position operators of the open string
endpoint. The infinite set of algebraic relations for the matrix elements of of
$\X{}$  was obtained. The Fock space representation of these infinite matrices
in terms of the operator expansion was found.

The expansion coefficients are recurrently related to each other, which allows
them to be explicitly computed in QCD2, and strongly reduced in general case.
The problem of elimination of unknown parameters in the Taylor expansion in
general case remains opened.

There is nothing like the  functional integral over random surfaces, which is
not so bad after all. Such integrals have problems even in critical dimension
of space, where there are tachyons. As for the noncritical strings, they do not
yet exist. Hopefully we would see such string theories in future\ct{Polyakov},
then they would have a chance to be equivalent to QCD.

The advantage of the present "string" theory is that it is manifestly
parametric invariant. We do not reduce the larger state to the parametric
invariant sector by Virasoro constraints, we directly work in physical sector.
In this sense we are building our "string model" as one dimensional topological
field theory.

We bypassed the hard unsolved problem of computation of the Wilson loops in
coordinate space. Not only these loops are hard to compute, they are not
defined as numbers in a usual sense, but rather as {\em distributions } in loop
space. The momentum loops which we compute instead, are expected to be the
usual numbers, with smooth momentum dependence beyond  perturbation theory.

Once again the large $N$ matrix technology proved its power. Our matrix model
is manifestly $O(d)$ invariant in any space dimension. Unlike the Eguchi-Kawai
model\ct{EK}, it has no matrix integrals, but rather algebraic equations. In a
way, this is simpler, as we could use variational methods with some trial
functions, dictated by physical intuition.

The Fock space of our theory is remarkably small. In fact, this is the very
similar to we would get in the Eguchi-Kawai model. This was the model of $d =
4$ random matrices with some invariant probability measure. Treating this model
by means of the operator expansion of the noncommutative probability theory we
would get the similar Fock space. In four dimensions this a space of all
sentences made from the four letter words -- an infinite space, but still a
discrete one.

There is a significant difference though. The EK model was the lattice model,
with inevitable large $N$ phase transition. The operator expansion in this
model would correspond to the strong coupling expansion of original lattice
gauge theory. Here, on the contrary, {the local limit is already taken}. We are
dealing with continuum theory, and we are expanding the Lorentz invariant
amplitudes in the low energy-momentum limit.

In terms of the Wilson loops, these amplitudes are given by path integrals over
all paths, with Fourier measure (not the Wiener measure!). The absolute
convergence of these integrals is to be achieved only at the nonperturbative
level, due to the area law. At small momenta, the dominant paths would cover
the area of the order of inverse string tension, which would lead to the
momentum expansion coefficients proportional to inverse powers of the string
tension.

These are not the directly observable quantities, because we do not include the
quark degrees of freedom at the endpoints of our string. Those are included
later  in the meson Hamiltonian we construct. There are two terms, the free
particle one, plus the contribution from the master field. This field is given
by the gradient of the delta function of the $\X{}$ operator. The matrix
elements of such master field between the trial states generated by the vertex
operators are well defined and calculable.

In the glueball sector we derived the linear wave equation which is supposed to
describe the mass spectrum. Again, the coefficients of this equation are
directly related to the coefficients of the operator expansion. Truncating this
equation would result in the finite linear equation in the glueball sector. In
a way, this is even simpler than the meson wave  equation.

We did not carry through these calculations, bat rather sketched the method.
Working out the details and deriving these wave equations would be the most
urgent next step.

An obvious question is how this agrees with the 't Hooft solution \ct{Hooft} in
two dimensions. I do not know the full answer to this question, but our meson
Hamiltonian  in the light cone gauge $ t = x_+ $ could be compared to the 't
Hooft's one. Clearly, the interaction term in the 't Hooft's Hamiltonian
corresponds to the string momentum operator $ P_- $ which is related to our
position field as discussed above.

So, we could interpret the 't Hooft solution in our terms. The real question is
how to get this solution from our equations. This would give us an insight for
the solution of the 4D problem. The momentum loop equation can be studied in
the light cone gauge, where all the dynamics reduces to Coulomb "gluons" with
propagators $  \inv{p_-^2} $.

The most appropriate would be the mixed $x_+, p_- $ representation, where  the
momentum loop equation must reproduce the Bethe-Salpeter equation for the
ladder diagrams. As was pointed out by 't Hooft, the $x_+, p_- $ propagation
goes along the  strip with fixed width in $p_-$ direction. It would be an
interesting exercise to push forward this analogy.

Regardless the 2D tests, the suggested approach  could be tested in 4D by
recovering the planar QCD graphs. The formal perturbation expansion follows
from direct iterations of the momentum loop equation\ct{Mig86}, assuming the
bare vacuum as a zeroth approximation.

Moreover, the perturbative QCD with running coupling must be recoverable in our
representation at large $p$ , or small $\X{}$. This must be an interesting
computation, making use of some WKB asymptotics.

Finally, why not try to numerically compute the hadron spectrum, truncating the
operator expansion and using variational approach like the one in\ct{AMi88}?
This is a hard problem, but with the algebraic formulation we are getting
closer to its solution.

\section{Acknowledgments}

I am grateful to  Robbert Dijkgraaf, Mike Douglas, David Gross, Yan Kogan, Ivan
Kostov, Andrei Matytsin, Greg Moore, Sasha Polyakov, Boris Rusakov, Misha
Shifman, Kumran Vafa,  Igor Veisburd and Ed Witten  for  interesting
discussions. This work was partially supported by the National Science
Foundation under contract PHYS-90-21984.

\appendix

\section{Functional Fourier Transform}

Let us parametrize the loops by the angle $ \theta $ which varies from $ 0 $ to
$2\pi$. The functional measure for the path is defined according to the
scalar product
\begin{equation}
  (A,B) = \oint \frac{d \theta}{2 \pi} A(\theta) B(\theta)
\end{equation}
which diagonalizes in the discrete Fourier representation (not to be mixed with
functional Fourier transform!)
\begin{equation}
  A(\theta) = \sum_{-\infty}^{+\infty} A_n e^{\imath n \theta}
\\;\;A_{-n} = A_n^{\star}
\end{equation}
\begin{equation}
  (A,B) =  \sum_{-\infty}^{+\infty} A_n B_{-n} =
A_0 B_0 + \sum_{1}^{\infty} a'_n b'_n + a''_n b''_n\\;\;
a'_n = \sqrt{2} \Re A_n,a''_n = \sqrt{2} \Im A_n
\end{equation}

The corresponding measure is given by an infinite product of the
Euclidean measures for the imaginary and real parts of each Fourier
component
\begin{equation}
  D Q = d^d Q_0 \prod_{1}^{\infty} d^d q'_n d^d q''_n
 \end{equation}
The orthogonality of Fourier transformation could now be explicitly
checked, as
\begin{eqnarray}
  \lefteqn{\int DC \exp \left( \imath \int d \theta C_{\alpha}(\theta)
	\left(
	 A_{\alpha}(\theta) - B_{\alpha}(\theta)
	\right) \right)
	}\\ \nonumber
  &=& \int d^d C_0 \prod_{1}^{\infty} d^d c'_n d^d c''_n \exp
\left( 2 \pi \imath
	\left(
	 C_0 \left(A_0-B_0 \right) +
	\sum_{1}^{\infty} c'_n\left(a'_n - b'_n \right)+
	 c''_n\left(a''_n - b''_n \right)
	\right)
\right)\\ \nonumber
&=&\delta^d\left(A_0-B_0 \right)
\prod_{1}^{\infty} \delta^d\left(a'_n - b'_n \right)
\delta^d\left(a''_n - b''_n \right)
\end{eqnarray}

Let us now check the parametric invariance
\begin{equation}
  \theta \rightarrow f(\theta)\\;\; f(2\pi) -f(0)
= 2\pi \\;\; f'(\theta) >0
\end{equation}
The functions $ x(\theta) $ and $ p(\theta) $ have
zero dimension in a sense, that only their argument transforms
\begin{equation}
  x(\theta) \rightarrow x \left( f(\theta) \right) \\;\;
 p(\theta) \rightarrow p\left( f(\theta) \right)
\end{equation}
The functions $ x'(\theta) $ and $ p'(\theta) $
have dimension $1$
\begin{equation}
  p'(\theta) \rightarrow  f'(\theta) p'\left( f(\theta) \right)\\;\;
x'(\theta) \rightarrow  f'(\theta) x' \left( f(\theta) \right)
\end{equation}

The invariance of the measure is easy to check for infinitesimal
reparametrization
\begin{equation}
  f(\theta) = \theta + \epsilon(\theta)\\;\; \epsilon(2\pi) = \epsilon(0)
\end{equation}
which changes $x$ and $(x,x)$ as follows
\begin{equation}
  \delta x(\theta) = \epsilon(\theta) x'(\theta) \\;\;
 \delta (x,x) = \oint \frac{d \theta}{2\pi}
\epsilon(\theta) 2 x_{\alpha}(\theta) x'_{\alpha}(\theta) =
-\oint \frac{d \theta}{2\pi}\epsilon'(\theta)x_{\alpha}^2(\theta)
\end{equation}
The corresponding Jacobian reduces to
\begin{equation}
  1 - \oint d \theta \epsilon'(\theta) =1
\end{equation}
in virtue of periodicity.

The computation of the functional determinants can be performed in terms of
discrete Fourier harmonics
\bea
J = \int Dx \delta^d(x(0)) \EXP{ \oh \i F_{\mu\nu} \int d\theta x_\mu(\theta)
x'_\nu(\theta)} = \br
\prod_{n=1}^{\infty} \int d^d x'_n d^d x''_n \EXP{ 2 \pi \i \, n\, F_{\mu\nu}
\lrb{x'_n}_\mu \lrb{x''_n}_\nu}.
\eea
This yields the infinite product
\be
J = \prod_{n=1}^{\infty} \lrb{n^d \, \left | \det F \right|}^{-1},
\ee
which we define by means of the $\zeta$ regularization
\be
J = \lim_{\alpha \ra 0}\,\EXP{- \sum_{n=1}^\8 \ln \lrb{n^d \, \left | \det F
\right|} \,\lrb{ n^d \, \left | \det F \right|}^{-\alpha}}
\ee
This yields
\be
J = const \, \left | \det F \right|^{-\zeta(0)} = const \sqrt{\left | \det F
\right|}
\ee

In odd dimensions this is zero, while in even dimensions it gives the pfaffian
of $F$.

The easiest way to obtain this result would be to note, that it comes from
elimination of the zero mode.
In the original form we could have locally rotated variables
\be
x_\mu(\theta) \Ra \Omega_{\mu\nu} x_\nu(\theta),
\ee
to reduce the antisymmetric matrix to a canonical Jordan form
\be
\oh \oint F_{\mu\nu} x_\mu d x_\nu \Ra \sum_{k=1}^{\oh d} f_k\oint p_k d q_k
\ee
Then we could rescale the coordinates
\be
p_k \Ra \frac{p_k}{\sqrt{f_k}},q_k \Ra \frac{q_k}{\sqrt{f_k}}
\ee
after which the matrix $F$ would disappear from the exponential. Only the delta
function  would acquire the corresponding Jacobian
\be
\delta(x(0)) \Ra \prod_{k=1}^{\oh d} f_k \delta(x(0)) = \sqrt{\det F}
\delta(x(0)).
\ee
This is the correct factor, so the remaining jacobian from rescaling of the
variables in the measure should cancel.

This is, indeed, so. Formally, the jacobian of this transformation is
\be
\lrb{\det \sqrt{F}}^{- \delta(0)}.
\ee
where $ \delta(\theta) $ is the periodic delta function
\be
\delta(\theta) = \sum_{-\8}^{\8} \EXP{\i n \theta} = 1 + 2 \sum_{1}^{\8} \cos(n
\theta).
\ee

Its limit at $\theta \ra 0 $ can be defined by analytic regularization
\be
\delta_{\alpha}(\theta) =  1 + 2 \sum_{1}^{\8} n^{-\alpha} \,\cos(n \theta).
\ee
We find
\be
\delta_{\alpha}(0) = 1 + 2\zeta(\alpha)
\ee
which in the limit $ \alpha \ra 0 $ yields zero, as it should.
This result means, that our measure with analytic regularization is scale
invariant $ x(\theta) \Ra \lambda x(\theta) $ in addition to parametric
invariance $ x(\theta) \Ra x(f(\theta)) $.

\section{Numerical Approach}

For the numerical solution these relations could be globally fitted by some
variational Anzatz for $\X{\mu}$ with finite large $N$, like it was done
before\ct{AMi88} for the discrete version of the momentum loop equation.

The discretization used in that work also corresponded to a polygon in momentum
space, but the discrete equations were valid only up to the powerlike
corrections at large number $L$ of vertexes of this polygon. The influence of
this discretization on the renormalizability of QCD is yet to be studied
theoretically. It could be that any approximation of this kind to the singular
integral equations destroys asymptotic freedom.

The  trial function was the Gaussian,
\bea
M\lrb{k_1,\dots k_L} = \int Dx  \,\EXP{ - \oq \int_0^T \dot{x}^2}
\,W(x)\,\EXP{- \i \sum k_l x(t_l)} \br
\approx Z_L \EXP{-\sum_{\omega = 2\pi n/T} |p_\omega|^2 Q(\omega)}.
\eea
The discrete Fourier expansion at the loop was used.
\be
p(t) = \sum_{l=1}^{L} k_l \theta(t-t_l) = \inv{T} \sum_{\omega = 2\pi n/T}
e^{\i\omega t} p_\omega, \; p_{-\omega} = p_\omega^{\star}, p_0 = 0.
\ee
The  variational function $Q(\omega)$ was parametrized by  eight parameters
(see\ct{AMi88}).

The Regge slope in this model is given by
\be
\alpha' = \frac{Q'(0)}{2\pi}.
\ee
The discrepancy of the momentum loop equation was squared, analytically
integrated over all momenta (with the Gaussian cutoff $\EXP{- \sum_{t=1}^{L}
\frac{p^2(t)}{\Lambda^2}}$ ) and then minimized for $ L = 3,4,...$. The
numerical limit $ L = \8 $ was reached at $ L \sim 40 $.

Numerically, the renormalization group behaviour
\be
\Lambda^2 \alpha' \propto \EXP{ \frac{48 \pi^2}{11 N g^2} + \frac{102}{121}
\log Ng^2},
\ee
was shown to agree with the variational solution of the discretized momentum
loop equation, but the accuracy in equation (about 0.1\%) was still not
sufficient to {\em derive} renormalization group behaviour numerically
(see\ct{AMi88} for more details).

The important advantage of the present set of equations for the dual amplitudes
is that {\em no approximations were made}. These are exact equations,
therefore, the variational solution with the general enough Anzatz must
reproduce asymptotic freedom. For example, one may try the above WKB
approximation with adjustable running coupling.


\begin{thebibliography}{99}

\bibitem{Hooft} A Planar Diagram Theory for Strong Interactions, by G. 't
Hooft, {\it Nucl. Phys. } B72 (1974) 461,
A Two-Dimensional Model for Mesons, by G. 't Hooft, {\it Nucl. Phys.} B75
(1974) 461.
\bibitem{MM} Yu.M.Makeenko, A.A.Migdal, Exact equation for the loop average in
multicolor QCD, {\it Phys. Lett.}, 88B, 135-137, (1979)
\bibitem{Mig83} Loop Equations and $\inv{N} $ Expansion, by A.A.Migdal,
{\it Physics Reports}, 102, 199-290, 1983,
\bibitem{KK} V.A.Kazakov and I.K.Kostov, Nucl. Phys. {\bf B176}
(1980) 199; Phys. Lett. {\bf B105} (1981) 453;
V.A.Kazakov, Nucl. Phys. {\bf B179} (1981) 283.
\bibitem{Witten} Baryons in $\inv{N}$ Expansion, by E. Witten, {\it Nucl.
Phys.} B160 (1979) 57.
\bibitem{EK} Reduction of Dynamical Degrees of Freedom in the Large-$N$ Gauge
Theory, by T. Eguchi and H. Kawai, {\it Phys. Rev. Lett.} 48 (1982) 1063.
\bibitem{Quench} The Quenched Eguchi-Kawai Model, by G. Bhanot, U.M. Heller and
H. Neuberger, {\it Phys. Lett.} B113 (1982) 47,
D. J. Gross and Y. Kitazawa, {\it Nucl. Phys.} B206 (1982)  440.
\bibitem{Cvitanovic}
CLASSICS ILLUSTRATED: GROUP THEORY. PART 1.
By Predrag Cvitanovic (Nordita), Print-84-0261 (NORDITA), Jan 1984. 206pp.
Nordita Notes,
THE PLANAR SECTOR OF FIELD THEORIES.
By Predrag Cvitanovic (Nordita), P.G. Lauwers (NIKHEF, Amsterdam), P.N.
Scharbach (Rutherford), NIKHEF-H/82-2, Jan 1982. 40pp.
Published in Nucl.Phys.B203:385,1982.
\bibitem{Voiculesku} D. V. Voiculescu, K. J. Dykema and A. Nica, {\it Free
Random Variables}, AMS 1992.
\bibitem{Douglas94} M.Douglas, Rutgers preprint, hep-th/9409098, Sept 1994.
\bibitem{Gross94} Rajesh Gopakumar, David Gross, {\it Mastering the Master
Field} PUPT 1520, Oct 1994.
\bibitem{Mig86} Momentum Loop Dynamics and Random Surfaces in QCD, by
A.A.Migdal, {\it Nucl. Phys.} B265 [FS15], 594-614, (1986),
\bibitem{BVM} Fourier Functional Transformations and Loop Equation. By
M.A. Bershadski, I.D. Vaisburd, A.A. Migdal (Moscow, Cybernetics Council),
1986. Yad. Fiz. 43 ( 1986) 503-513.
\bibitem{Pol87}  Gauge Fields and Strings, by A.M.Polyakov, {\it Harwood
Academic Publishers} , 1987.
\bibitem{AMi88} Variational Solution of the Loop Equation in QCD, by
M.E.~Agishtein and A.A.~Migdal, Nucl.Phys.B(Proc. Suppl.) 4 (1988) 59-63.
\bibitem{Polyakov} A.M. Polyakov (to be publiushed).
\end{thebibliography}
\end{document}